\documentclass[ reprint, superscriptaddress,
bibnotes, amsmath,amssymb, aps, prl
pra, prb, rmp, prstab, prstper, floatfix,
]{revtex4-2}

\usepackage{graphicx}
\usepackage{dcolumn}
\usepackage{bm}
\usepackage[super]{nth} 
\usepackage{dsfont} 

\usepackage[hidelinks]{hyperref}

\usepackage{orcidlink} 
\usepackage{mathtools} 
\usepackage{cleveref} 
\usepackage{xcolor} 

\usepackage{subcaption}

\newcommand{\ket}[1]{\left\vert#1\right\rangle}
\newcommand{\bra}[1]{\left\langle#1\right\vert}

\newcommand{\pTrace}[2]{\underset{#1}{Tr}\left[#2\right]}

\newcommand\redsout{\bgroup\markoverwith{\textcolor{red}{\rule[0.5ex]{2pt}{0.4pt}}}\ULon}

\newcommand{\fs}[3]{f_{#1}^{#2}(t_{#3})}

\definecolor{ppblue}{RGB}{46,117,182}
\definecolor{ppred}{RGB}{197, 90, 17}

\newcommand{\HH}[0]{\hat{H}}

\begin{document}
\preprint{}
\title{Memory-Induced Limits on Entanglement Distribution}

\author{Roberto Benjamin Salazar Vargas\,\orcidlink{0000-0003-1737-1433}}
\email{roberto.salazar@um.edu.mt}
\affiliation{Department of Communications and Computer Engineering, University of Malta, Msida MSD 2080, Malta}

\author{Hayden Zammit\,\orcidlink{0009-0003-9917-929X}}
\affiliation{Department of Physics, University of Malta, Msida MSD 2080, Malta}

\author{Trevor Spiteri\,\orcidlink{0000-0002-7954-5534}}
\affiliation{Department of Communications and Computer Engineering, University of Malta, Msida MSD 2080, Malta}

\author{Gianluca Valentino\,\orcidlink{0000-0003-3864-7785}}
\affiliation{Department of Communications and Computer Engineering, University of Malta, Msida MSD 2080, Malta}

\author{Tony J. G. Apollaro\,\orcidlink{0000-0002-9324-9336}}
\affiliation{Department of Physics, University of Malta, Msida MSD 2080, Malta}

\author{Johann A. Briffa\,\orcidlink{0000-0003-3532-9474}}
\affiliation{Department of Communications and Computer Engineering, University of Malta, Msida MSD 2080, Malta}

\date{\today}

\begin{abstract}
   We investigate the operational limits of entanglement distribution through quantum channels reused without reinitialization. Building on an effective repeated-use channel description, we show that residual excitations left in the channel from previous uses act as a memory source that progressively suppresses coherent quantum transmission. To quantify this limitation, we derive an analytical lower bound on the quantum capacity and combine it with the Holevo–Werner semidefinite-programming upper bound. We also provide an analytical criterion to identify entanglement-breaking channels, with thresholds that depend jointly on chain length and reuse. For perfect-state-transfer chains, the resulting capacity window yields tighter estimates than previous bounds, with the relative improvement becoming substantially more pronounced closer to the entanglement-breaking threshold.  Moreover, on the second use, our lower bound certifies a maximal relative timing-error tolerance across all chains with internal memory that lies just below 22.63\% and is attained by the three-site chain.  We further derive a fundamental relative timing-error threshold of 36.41\%, above which we analytically guarantee zero unassisted quantum capacity for every chain with internal memory, independent of the number of uses. These results establish a clear benchmark model for repeated entanglement distribution, and identify timing precision, chain length, and memory accumulation as fundamental constraints on scalable quantum-network architectures.
\end{abstract}

\maketitle

\section{Introduction}
Entanglement distribution is a central primitive for quantum communication and networked quantum information processing \cite{Horodecki2009Review,Perseguers2013,Kimble2008,Wehner2018}. Establishing entanglement between spatially separated nodes underlies quantum teleportation, entanglement-based quantum key distribution, distributed quantum computing, and modular quantum architectures \cite{Bennett1993,Ekert1991,Monroe2014,Barral2025}. Identifying the mechanisms that limit repeated entanglement distribution is therefore essential for assessing the scalability of quantum networks \cite{Perseguers2013,Duan2001,Sangouard2011}.

A standard route to entanglement distribution is quantum state transfer \cite{Bose2003, Nikolopoulos2014, 10.1063/1.4978327, apollaroEntangledStatesAre2022}. If two qubits are initially entangled locally, entanglement can be distributed by transferring one member of the pair through a quantum channel while retaining the other as an ancilla at the sender node \cite{Bose2003,Bose01012007}. The final entanglement between the ancilla and the remote receiver depends on how well the channel preserves the coherence of the transmitted system. Thus, quantum state transfer provides an elementary dynamical building block for entanglement distribution.

A common idealization in quantum-state-transfer protocols is that the channel is reset to a fixed reference state before each use \cite{navascuesResettingUncontrolledQuantum2018,Reed2010,Magnard2018,bassmanoftelieDynamicCoolingContemporary2024}. Then each transmission is described by the same memoryless channel, whose performance can be characterized by single-use quantities such as the average transfer fidelity, concurrence, or coherent information \cite{Bose2003,Lorenzo2022,apollaroEntangledStatesAre2022,Barnum1998}. In realistic architectures, however,  resetting a quantum channel to its initial state after each transmission may incur additional temporal, control, and energetic costs \cite{navascuesResettingUncontrolledQuantum2018,Reed2010,Magnard2018,
bassmanoftelieDynamicCoolingContemporary2024,Sur2026}, while precise transfer timing constitutes a further control requirement \cite{xuerebImpactImperfectTimekeeping2023,meierPrecisionNotLimited2025b}. On the other hand, if the same physical channel is reused without full reinitialization, residual excitations left by previous transmissions may affect subsequent ones \cite{navascuesResettingUncontrolledQuantum2018,bassmanoftelieDynamicCoolingContemporary2024}. The effective channel seen by a given input then depends on the history of previous uses \cite{carusoQuantumChannelsMemory2014b}.

Memory effects, therefore, represent a potential threat to entanglement distribution \cite{PhysRevA.69.012306, Giovannetti_2005, bayatMemoryEffectsSpinchain2008a}. Unlike fixed single-use imperfections, memory effects accumulate dynamically. Even a channel engineered for perfect state transfer at an ideal readout time can, under imperfect temporal control, become a memory-bearing medium with enhanced dephasing and reduced coherent transmission \cite{bayatMemoryEffectsSpinchain2008a,xuerebImpactImperfectTimekeeping2023,Chen2019}. Indeed, any timing mismatch in a single use can leave the population in the channel's internal degrees of freedom, which then serves as an effective environment for later transmissions \cite{bayatMemoryEffectsSpinchain2008a,PhysRevA.69.012306,Giovannetti_2005,Giovannetti2009}. Other scenarios in which memory effects arise  during successive uses of a quantum channel without reinitialization include channels that are not engineered for perfect state transfer~\cite{Bose2003, apollaro99FidelityBallistic2012,Bezaz_2025}.

This setting has been recently analyzed for repeated quantum-state transfer via spin-chain channels \cite{bayatMemoryEffectsSpinchain2008a,Giovannetti2009,Zammit:2025orr}. For U(1)-symmetric channels generated by quadratic fermionic Hamiltonians, the repeated-use map can be written analytically in terms of only single-particle transition amplitudes \cite{Zammit:2025orr}. Previous uses enter through memory factors that modify the average transfer fidelity and, through the state-transfer picture above, the channel's ability to distribute entanglement \cite{Chen2019,Zammit:2025orr}.  Equal readout timing errors at each channel use lead to a systematic reduction in transfer fidelity and a shrinking time window for successful entanglement distribution \cite{Zammit:2025orr}. The same framework yields an effective decomposition into generalized amplitude-damping and dephasing contributions, explicitly connecting channel memory with noise accumulation.

While these results establish the role of memory effects on specific quantum information processing tasks, e.g., quantum state transfer, a sharper operational characterization requires bounds on the quantum capacity of the effective channel associated with each reuse. The quantum capacity gives the asymptotic rate at which quantum information, and hence entanglement, can be reliably transmitted through a channel \cite{Lloyd1997, Devetak2005, Barnum1998}. It is therefore a natural benchmark for determining whether repeated channel use remains useful for scalable entanglement distribution. Capacity bounds can distinguish regimes in which reduced fidelity still permits quantum communication from regimes in which memory-induced dephasing suppresses it \cite{Khatri:2020dfc,Wilde2017}.

In this work, we develop tighter estimates of the quantum capacity for memory-affected repeated-use channels. Building on the effective channel model introduced above, we derive an analytical lower bound on the quantum capacity and combine it with the Holevo--Werner semidefinite-programming upper bound \cite{HolevoWerner2001,FangFawzi2021}. Additionally, we derive an analytical positive-partial-transpose criterion for entanglement breaking, with thresholds that depend jointly on chain length and reuse. The resulting capacity window improves on previous estimates, with the relative improvement becoming substantially more pronounced closer to the entanglement-breaking threshold. 

We apply these methods to quantify the effect of readout timing errors in a perfect-state-transfer spin chain. We derive chain-length-dependent zero-capacity thresholds and show that relative timing errors of 36.41\% or higher guarantee zero unassisted quantum capacity for every chain with internal memory, independently of the number of uses. This guarantee follows from damping-induced antidegradability. We also analytically establish that, on the second use, the maximal relative timing-error tolerance certified by our lower bound across all chains with internal memory lies just below 22.63\% and is attained by the three-site chain. These thresholds guide our analysis of representative timing-error regimes. For small timing errors, our lower bound certifies positive capacity over several uses, while memory accumulation progressively suppresses the certified transmission rates. The positive-partial-transpose criterion further identifies entanglement-breaking regimes below the universal zero-capacity threshold. Together, these results establish a benchmark for repeated entanglement distribution and quantify the constraints that timing precision, chain length, and memory accumulation impose on scalable quantum-network architectures.

The paper is organized as follows. We first review the repeated-use channel model and its effective dynamical-map representation. We then derive the analytical lower bound on the quantum capacity and describe the Holevo–Werner semidefinite-programming upper bound used as a converse estimate. Additionally, we derive an analytical positive-partial-transpose criterion to identify entanglement-breaking regimes. We next apply these tools to the repeated-use spin-chain channel with relative readout timing errors, first deriving chain-length-dependent and universal thresholds for zero unassisted quantum capacity and identifying the maximal timing-error tolerance certified by the lower bound on the second use. These thresholds guide the subsequent analysis of representative error regimes, where we examine how the capacity window and entanglement-breaking transitions depend on chain length, the number of uses, and the relative readout timing error. Finally, we discuss the implications of these results for entanglement distribution in scalable quantum-network architectures, emphasizing timing precision, memory accumulation, and channel reset as operational constraints.

\section{Repeated-use channel model}

We first derive the effective channel associated with a single
state-transfer operation. We then extend the construction to
sequential transfers through the same spin chain without
reinitializing its internal sites. This allows us to identify
separately the population-transfer loss and the memory-induced
suppression of coherence.

\subsection{Single-Use State Transfer}
In single-use quantum-state transfer through a spin-$1/2$ chain, the goal is to transfer a pure state $|\psi\rangle_1=a|0\rangle+b|1\rangle$ from one end of the chain to the other. Formally this implies evolving the total system from the initial state
\begin{equation}
    \ket{\psi}_{1\cdots N}=\ket{\psi}_{1}\otimes\ket{0\cdots0}_{2\cdots N}
\end{equation}
to the final state
\begin{equation}
    \ket{\psi(t)} = e^{-i\HH t}\ket{\psi} = \alpha\ket{0} + \beta e^{-i\HH t}\ket{1} = \alpha\ket{0} + \beta \ket{N}
\end{equation}
as closely as possible. For any time $t$, the state is given by
\begin{equation}
    \ket{\psi(t)} = \alpha\ket{0} + \beta \sum_{i=1}^{N}f_{1}^{i}(t)\ket{i}
\end{equation}
in terms of transition amplitudes $f_{i}^{j}(t)=\bra{j}U(t)\ket{i}$ from the $i$th to the $j$th site. Here, we denote by $|j\rangle\equiv|00\ldots 1_j00\ldots\rangle$ the state in which the spin at site $j$ is flipped. In terms of these amplitudes, the generally mixed state of the final site is: 
\begin{align}\label{eq_final_evolved_reciever}
    &\hat{\rho}_{N}(t)=\pTrace{1,2,\dots,N-1}{e^{-i\HH t}\ket{\psi}\bra{\psi}e^{i\HH t}}\\
    &= \begin{pmatrix}
    1-|\beta|^{2}|f_{1}^{N}(t)|^{2} & \alpha\beta^{*} f_{1}^{N}(t)^{*} \\
    \alpha^{*}\beta f_{1}^{N}(t) & |\beta|^{2}|f_{1}^{N}(t)|^{2}
    \end{pmatrix}~,
\end{align}
which can be expressed as
\begin{equation}\label{eq_dyn_map_1}
    \begin{pmatrix}
        \rho_{00}\\
        \rho_{10}\\
        \rho_{01}\\
        \rho_{11}
    \end{pmatrix}_{N} \!\!\!\!\!\!= \!
    \begin{pmatrix}
        1 & 0 & 0 & 1-|\fs{1}{N}{}|^{2}\\
        0 & \fs{1}{N}{}^{*} & 0 & 0\\
        0 & 0 & \fs{1}{N}{} & 0\\
        0 & 0 & 0 & |\fs{1}{N}{}|^{2}
    \end{pmatrix}\!\!
    \begin{pmatrix}
        |\alpha|^{2}\\ 
        \alpha\beta^{*}\\
        \alpha^{*}\beta\\
        |\beta|^{2}
    \end{pmatrix}_{1}~\!\!\!\!\!,
\end{equation}
in terms of the CPTP dynamical map $\Phi$~\cite{Society2004,Lorenzo2022} which satisfies $\hat{\rho}_{N}(t) = \Phi(\hat{\rho}_{1})$. Equation (\ref{eq_dyn_map_1}) is an amplitude-damping channel with damping probability $ 1-|f_{1}^{N}(t)|^{2}$. Thus, for a single use, transfer loss is completely determined by the end-to-end transition amplitude.

\subsection{Repeated-Use State Transfer}

\begin{figure*}[t]
    \centering
    \includegraphics[width=0.72\textwidth]{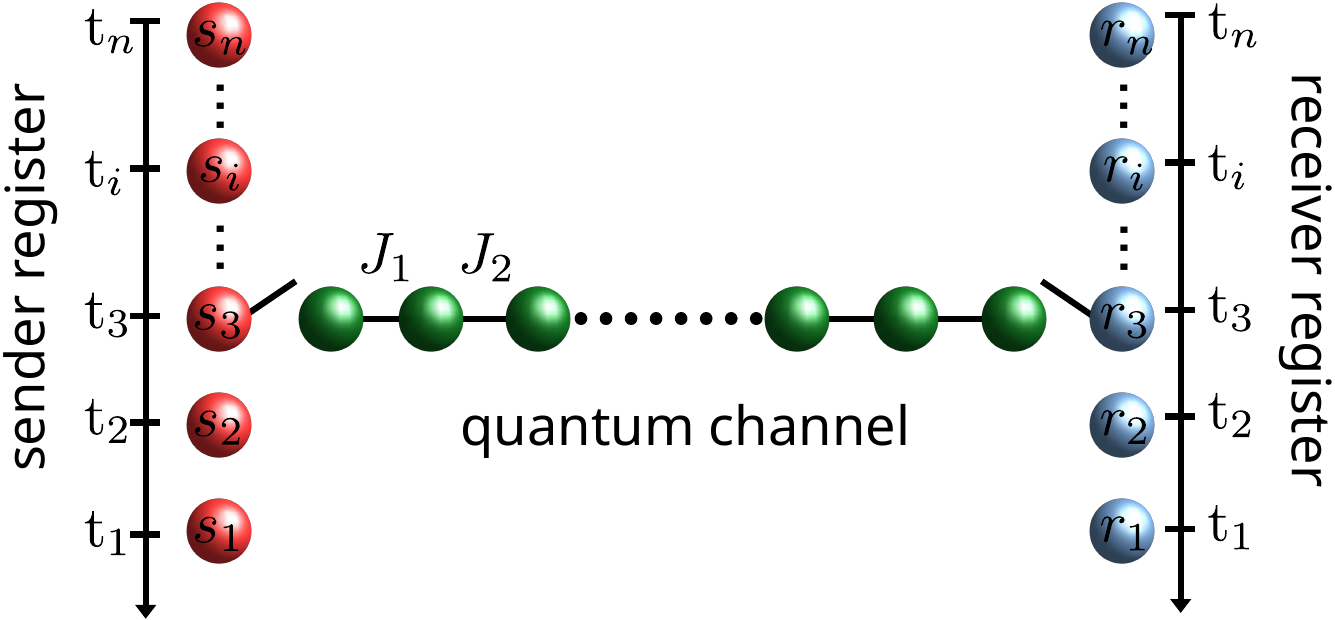}
    \caption{Protocol for transferring quantum information from an
    $n$-qubit sender register (red spheres) to an $n$-qubit receiver
    register (blue spheres) through repeated use of the same quantum
    channel (green spheres). The $i$th transfer occurs at time $t_i$.}
    \label{fig_model}
\end{figure*}

The goal of the repeated-use protocol is to transfer a register of $n$ pure states to a receiver register initialized in $|0\rangle^{\otimes n}$, using the same spin chain for each transfer without reinitializing its internal sites to the fully polarized state $|0\cdots0\rangle_{2\cdots N-1}$ after each use. We consider a system composed of $N$ sites, with the first, last, and intermediate sites denoted as the sender, receiver, and chain, respectively. For $N=2$, the system contains only the sender and receiver sites and therefore has no internal degree of freedom that remains after a transfer. Since both endpoint sites are removed and replaced between successive uses in the protocol, no residual excitation can be retained from one transmission to the next. The two-site channel may still exhibit timing-induced transfer loss, but it cannot develop the accumulated internal memory studied here, and its effective map is consequently independent of the reuse number. We therefore restrict the repeated-use memory analysis to $N\geq 3$, for which at least one internal spin is retained between successive transfers.

The sequential use of the quantum channel is illustrated in Fig.~\ref{fig_model}. During use $k$,  the sender qubit is prepared in $\left|\psi_{k}\right\rangle =a_{k}\left|0\right\rangle +b_{k}\left|1\right\rangle $,
the receiver site is replaced by a fresh qubit in $\left|0\right\rangle $,
and the full system evolves for a time $t_{k}$. The sender and receiver
sites are then removed, whereas the spins in the chain retain the
state produced by the previous transfers which will impact all subsequent uses of the channel.

Consequently, the state of the receiver at the $n$th use then depends on all transfer times $t_{n;1} = (t_{1}, t_{2},\dots,t_{n})$ and on all the amplitudes $a_{n;1},b_{n;1}$ that define the pure states of the previous sender states:
\begin{equation}
    \hat{\rho}_{R}^{(n)}(a_{n;1},b_{n;1}|t_{n;1}) = \pTrace{S,C}{U(t_{n}) \hat{\rho}_{SCR}U^{\dagger}(t_{n})}~,
\end{equation}
where
\begin{equation}
    \rho_{SCR}=\hat{\rho}_{S}^{(n)}(a_{n},b_{n})\otimes\hat{\rho}_{C}^{(n-1)}(a_{n-1;1},b_{n-1;1}|t_{n-1;1})\otimes\hat{\rho}_{R}
\end{equation} 
Using the above expression requires full knowledge of the $a_{n;1}$ and $b_{n;1}$ that define each use. When the transmitted states are treated as unknown, it is appropriate to consider the average effect of an imperfect channel use. As a consequence, the resulting effective map depends on the number of previous uses, but not on their particular input amplitudes. In this case, the average receiver state on the  $n$th use  $\overline{\rho}_{R}^{(n)}(t_{n;1})$ is given by
\begin{equation}
    \pTrace{S,C}{U(t_{n}) \left(\hat{\rho}_{S}^{(n)}(a_{n},b_{n})\overline{\rho}_{C}^{(n-1)}(t_{n-1;1})\hat{\rho}_{R}\right)U^{\dagger}(t_{n})}~,
\end{equation}
where $\overline{\rho}_{C}^{(n-1)}(t_{n-1;1})$ denotes the average $n^{\mathrm{th}}$ use channel, defined recursively as
\begin{align}
     & \overline{\rho}_C^{(n)}\left(t_{n;1}\right)=\pTrace{S,R}{\hat{\rho}_{SRC}\left(t_{n;1}\right)}                                                                           \\
     & =\int d\psi~\pTrace{S,R}{\hat{U}(t_{n})\hat{\rho}_{S}^{(n)}\overline{\rho}_{C}^{(n-1)}\left(t_{n-1;1}\right)\hat{\rho}_{R}^{(n)}\hat{U}^{\dagger}(t_{n})}~,
\end{align}
where the integral is over all pure input states.

It was shown previously in Ref \cite{Zammit:2025orr} that, for systems whose Hamiltonian is both $U(1)$-symmetric and quadratic in its fermionic representation, the dynamical map satisfying  $\rho_{R}^{(n)}= \Phi_{n}^{(N)}(\overline{\rho}_{S}^{(n)})$ is:
\begin{widetext}
    \begin{equation}\label{eq_nth_use_map}
\Phi_{n}^{(N)}\left(t_{n;1}\right)=\begin{pmatrix}
            1\!-\!\frac{1}{2}\!\displaystyle\sum_{i=1}^{n-1}\left|B_{i}^{(N)}(t_{n;n-i})\right|^{2} & 0  & 0  & 1\!-\!\frac{1}{2}\!\displaystyle\sum_{i=1}^{n-1}\left|B_{i}^{(N)}(t_{n;n-i})\right|^{2}\!-\!\left|\!f_{1}^{N}\!\left(t_{n}\right)\!\right|^{2} \\ 
            0  & f_{1}^{N}\!\left(t_{n}\right)^{\ast} A_{n-1}^{(N)}\left(t_{n-1;1}\right) & 0   & 0        \\
            0   & 0   & f_{1}^{N}\!\left(t_{n}\right) A_{n-1}^{(N)}\left(t_{n-1;1}\right) & 0   \\
            \frac{1}{2}\!\displaystyle\sum_{i=1}^{n-1}\left|B_{i}^{(N)}(t_{n;n-i})\right|^{2}       & 0   & 0 & \frac{1}{2}\!\displaystyle\sum_{i=1}^{n-1}\left|B_{i}^{(N)}(t_{n;n-i})\right|^{2}\!+\!\left|\!f_{1}^{N}\!\left(t_{n}\right)\!\right|^{2}
\end{pmatrix}~,
\end{equation}
\end{widetext}
We note that \cref{eq_nth_use_map} differs from \cref{eq_dyn_map_1} through the presence of the two memory terms $A_{n-1}^{(N)}\left(t_{n-1;1}\right)$ and $B_{i}^{(N)}(t_{n;n-i})$. Note that this map can be recast as the product between the superoperators
\begin{widetext}

\begin{equation}
 \begin{pmatrix}
1 & 0 & 0 & 0\\
0 & A_{n-1}^{(N)}(t_{n-1;1}) & 0 & 0\\
0 & 0 & A_{n-1}^{(N)}(t_{n-1;1}) & 0\\
0 & 0 & 0 & 1
\end{pmatrix}~,
\begin{pmatrix}
            1-\mathcal{B}_{n-1}^{(N)}
            & 0  & 0  & 1-\mathcal{B}_{n-1}^{(N)}-\left|f_{1}^{N}\left(t_{n}\right)\right|^{2} \\ 
            0  & f_{1}^{N}\left(t_{n}\right)^{\ast} & 0   & 0        \\
            0   & 0   & f_{1}^{N}\left(t_{n}\right) & 0   \\
            \mathcal{B}_{n-1}^{(N)}       & 0   & 0 & \mathcal{B}_{n-1}^{(N)}+\left|f_{1}^{N}\left(t_{n}\right)\right|^{2}
\end{pmatrix}~,
\end{equation} 
\end{widetext}
where $\mathcal{B}_{n-1}^{(N)}\equiv\frac{1}{2}\!\displaystyle\sum_{i=1}^{n-1}\left|B_{i}^{(N)}(t_{n;n-i})\right|^{2}$.

Throughout, $N$ denotes the chain length and $n$ the channel-use number. When comparing different chain lengths, we make the dependence on $N$ explicit through a parenthesized superscript, as in $\Phi_n^{(N)}$; otherwise, this dependence is left implicit. We similarly display dependence on the readout times or relative timing error when needed.

Since, $\mathcal{B}_{n-1}^{(N)},A_{n-1}^{(N)}\in [0,1]$, these two operators can be associated with dephasing (PD) and generalized-amplitude-damping (GAD) channels, respectively \cite{Khatri2020,Devetak2005,Puch2021}. Indeed, they can be brought into standard form by defining the dephasing parameter, damping parameter, and mixing parameter, according to $\lambda_{n}^{(N)}=1-(A_{n-1}^{(N)}(t_{n-1;1}))^{2}$, $\gamma_{n}^{(N)}=1-\left|f_{1}^{N}\left(t_{n}\right)\right|^{2}$, and $p_{n}^{(N)}=1-(\mathcal{B}_{n-1}^{(N)}/\gamma_{n}^{(N)})$, respectively.
These definitions yield the decomposition:
\begin{align}
\Phi_{n}^{(N)}(t_{n;1})=\Phi_{\mathrm{GAD}}\left(\gamma_{n}^{(N)},p_{n}^{(N)}\right)\Phi_{\mathrm{PD}}\left(\lambda_{n}^{(N)}\right)~,
\label{eq:decom}
\end{align}
which we use throughout our analysis. For fixed $n$, $\Phi_{n}^{(N)}$ denotes the averaged marginal map experienced by the nth input after $n-1$ previous transfers. Accordingly, the channel capacity $Q(\Phi_{n}^{(N)})$ characterizes independent realizations of this effective marginal map. For later use, we define the relative readout timing error as
\begin{equation}
    \delta=\frac{|t-\tau|}{\tau},
\end{equation}
where $\tau$ is the ideal readout time. Our work therefore quantifies the degradation of the nth-use communication link due to a readout timing error when the internal memory is inaccessible and collective coding across the physical sequence is excluded.

\section{Methods}
The effective channel obtained on the $n$th use has the form of a dephasing channel  followed by a generalized-amplitude-damping channel, as shown in Eq. (\ref{eq:decom}). This decomposition gives a direct physical interpretation of the memory effects: the parameter $\gamma_n$ captures the loss of population transfer, $p_n$ encodes the effective excitation imbalance induced by previous channel uses, and $\lambda_n$ quantifies the accumulated dephasing generated by the memory factor $A_{n-1}$. However, although the two components have well-understood limiting cases, their composition is not generally degradable (see Appendix A). Therefore, the coherent information is not known to be additive, and the quantum capacity of $\Phi_n$ cannot in general be evaluated from a single-letter formula.

\subsection{Quantum-capacity bounds}
The quantum capacity $Q(\Phi)$ is the asymptotic rate at which
quantum information can be transmitted reliably through a
channel $\Phi$. It is given by the regularized coherent
information \cite{NielsenChuang2010, Khatri:2020dfc, Barnum1998}:
\begin{equation}
    Q(\Phi)
    =
    \lim_{m\rightarrow\infty}
    \frac{1}{m}
    \max_{\rho}
    I_c\!\left(\rho,\Phi^{\otimes m}\right),
    \label{eq:quantum-capacity}
\end{equation}
where
\begin{equation}
    I_c(\rho,\Phi)
    =
    S\!\left(\Phi(\rho)\right)
    -
    S\!\left(\Phi^c(\rho)\right).
    \label{eq:coherent-information}
\end{equation}
Here, $S(\rho)=-\mathrm{Tr}(\rho\log\rho)$ denotes the von Neumann entropy, and $\Phi^{c}$ is the complementary channel. Additionally,
for a given Kraus representation of the dynamical map $\Phi$ in channel form $\mathcal{E}_\Phi$:
\begin{equation}
\mathcal{E}_\Phi\left(\rho\right)=\sum_{i=1}^{r}K_{i}\rho K_{i}^{\dagger},\quad\sum_{i=1}^{r}K_{i}^{\dagger}K_{i}=I.
\end{equation}
The corresponding complementary channel $\mathcal E_\Phi^c$ is: 
\begin{equation}
\mathcal{E}^{c}_\Phi\left(\rho\right)=\sum_{i,j=1}^{r}\mathrm{Tr}\left[K_{i}\rho K_{j}^{\dagger}\right]\left|i\right\rangle \!\left\langle j\right|,\label{eq:complchan}
\end{equation}
where $r$ denotes the Kraus rank of $\mathcal E_\Phi$. The complementary channel gives the environmental output in a Stinespring representation of $\mathcal E_\Phi$ and takes values in $\mathcal B(\mathcal H_E)$, with $\dim(\mathcal H_E)=r\leq\dim(\mathcal H_A)\dim(\mathcal H_B)$. We denote the associated dynamical map by $\Phi^c$.

For degradable channels, the coherent information is additive,
and the regularization reduces to
\begin{equation}\label{chandeg}
    Q(\Phi)=\max_\rho I_c(\rho,\Phi).
\end{equation}
The single-use amplitude-damping and dephasing channels are
degradable in the appropriate parameter regimes \cite{Devetak2005,Wolf2007}. Their
composition in Eq.~\eqref{eq:decom}, however,
is generally nondegradable, as shown in Appendix~A.
Consequently, its quantum capacity cannot in general be
obtained from a single-letter expression.

We therefore characterize the repeated-use channel through upper and lower bounds on its quantum capacity. The goal is not only to determine whether the channel remains capable of transmitting quantum information, but also to quantify how tightly this capability can be constrained as memory effects accumulate.

Previously, the following bottleneck upper bound \cite{Wolf2007} was applied to the dynamical map $\Phi_n$ in Ref. \cite{Zammit:2025orr},
\begin{equation}\label{eq:bottleneck inequality}
  \mathrm{BN}(n) = \mathrm{min}\left\{ Q(\Phi_{\mathrm{GAD}}\left(\gamma_{n},0\right)), Q(\Phi_{\mathrm{PD}}\left(\lambda_{n}\right))\right\}.
\end{equation}
which satisfies the inequality $ Q(\Phi_{n}(t_{n;1}))\leq\mathrm{BN}(n)$.

This upper bound is useful because it separates the limitations associated with damping and dephasing.  However, as a bottleneck bound, it may overestimate the capacity of the composite channel because it does not fully account for the combined action of amplitude damping and dephasing.

Accordingly, for each use $n$, the admissible capacity interval is:
\begin{equation}\label{capwind}
\mathrm{LB}(n)\leq Q\left(\Phi_{n}\left(t_{n:1}\right)\right)\leq\mathrm{HW}\left(n,l\right),
\end{equation}
where the lower bound is derived in Appendix B, while the upper bound
is a Holevo--Werner bound computed using an SDP optimization at the level $l$.
This capacity window provides the operational quantity used in Sec.
IV to assess the loss of entanglement-distribution capability under
repeated channel use. Explicitly, the analytical lower bound $\mathrm{LB}(n)$
is given by:

\begin{eqnarray}
\mathrm{LB}(n) & = & \max_{z\in\left[-1,1\right]}\left\{ h\!\left((1+(1-\gamma_{n})z+\gamma_{n}(2p_{n}-1))/2\right)\right.\nonumber \\
 & - & \left.\!H\!\left(\lambda_{+},\lambda_{-},p_{n}\gamma_{n}a_{-},(1-p_{n})\gamma_n a_{+}\right)\right\}, \label{eq:lowerboundZ}
\end{eqnarray}
where $h(x)$ is the binary entropy and $H(x)$ is the
Shannon entropy for the distribution of the four probabilities $\left\{\lambda_{+},\lambda_{-},p_{n}\gamma_{n}a_{-},(1-p_{n})\gamma_{n}a_{+}\right\}$,
with $a_{\pm}=(1\pm z)/2$ and:
\begin{widetext}
\begin{equation}
\lambda_{\pm}\!=\!\frac{1}{2}\left[\!\left(a_{+}b_{0}\!+\!a_{-}b_{1}\right)\!\pm\!\sqrt{\left(a_{+}b_{0}\!-\!a_{-}b_{1}\right)^{2}\!\!+\!4a_{+}a_{-}\!\left(1\!-\!\gamma_{n}\right)\!\!\left(1\!-\!\lambda_{n}\right)}\right]
\end{equation}
\end{widetext}
with $b_{0}=1-(1-p_{n})\gamma_{n}$, and $b_{1}=1-p_{n}\gamma_{n}$.

The Holevo--Werner upper bound $\mathrm{HW}(n,l)$ is obtained from an SDP that evaluates the geometric R\'{e}nyi channel divergence $R_{\alpha,\Theta}(\mathcal E_{\Phi_n})$ for $\alpha=1+2^{-l}$ with respect to the set of subchannels $\mathcal V_\Theta$ that preserve the positive-partial-transpose (PPT) property. More precisely, for
an arbitrary channel $\mathcal{N}$ \cite{HolevoWerner2001}: 
\begin{equation}
R_{\alpha,\Theta}\left(\mathcal{N}\right)=\min_{\mathcal{M}\in\mathcal{V}_{\Theta}}D_{\alpha}\left(\mathcal{N}\parallel\mathcal{M}\right)
\end{equation}
with $D_{\alpha}\left(\mathcal{N}\parallel\mathcal{M}\right)$ the
R\'{e}nyi channel divergence \cite{FangFawzi2021}. Due to the monotonicity of the R\'{e}nyi channel divergence for $\alpha>1$ this upper bound satisfies: 
\begin{equation}
Q\left(\Phi_{n}\left(t_{n:1}\right)\right)\leq\mathrm{HW}\left(n,l+1\right)\leq\mathrm{HW}\left(n,l\right)
\end{equation}
for every $l\in\mathbb{N}$. The full description of the SDP formulation of $\mathrm{HW}(n,l)$ is provided in Appendix C.

We remark that although the output states associated with a fixed effective map $\Phi_n$ may, in principle, be processed by entanglement-distillation protocols, such processing cannot overcome a vanishing Holevo–Werner bound. Indeed, the PPT-type bound applies to arbitrary blocklengths and therefore upper-bounds PPT-assisted, and hence LOCC-assisted, quantum communication. Therefore, when $R_{\alpha,\Theta}\left(\mathcal{E}_{\Phi_n}\right)=0$, the two-way-assisted quantum capacity also vanishes, ruling out asymptotic entanglement distillation at a nonzero rate. This conclusion concerns repeated independent realizations of the fixed map $\Phi_n$; it does not invoke coherent access to the spin-chain memory or to correlations between different effective maps.

Together, the bottleneck upper bound, the analytical lower bound, and the Holevo–Werner SDP bound provide complementary information. The lower bound certifies a rate achievable by the repeated-use channel; the bottleneck bound provides a physically transparent but generally loose converse estimate; and the Holevo–Werner hierarchy tightens this converse by incorporating the full structure of the composite channel. These bounds form the basis for the capacity-window analysis presented in the next section.

\subsection{Entanglement-breaking threshold}

Beyond constraining the capacity between upper and lower
bounds, the Choi state permits us to identify an exact
zero-capacity region. To determine precisely when
entanglement distribution becomes impossible, we apply the PPT
criterion to the Choi state of the effective channel $\Phi_n$.
For the channel parameters considered here, this condition reads
\begin{equation}
    (1-\gamma_n)(1-\lambda_n)
    \leq \gamma_n^2 p_n(1-p_n).
\end{equation}
Since $\Phi_n$ is a qubit-to-qubit channel, positivity of the
partially transposed Choi state implies its separability.
Consequently, $\Phi_n$ is entanglement breaking and therefore has zero entanglement distribution capacity. In light of this, it is useful to define the critical reuse number: 
\begin{equation}
    n_{\mathrm{EB}}(\delta)
    =
    \min\left\{
    n:\,
    (1-\gamma_n)(1-\lambda_n)
    \leq\gamma_n^2p_n(1-p_n)
    \right\}.
\end{equation}
where, by definition, the parameters $\gamma_n$, $\lambda_n$, and $p_n$ depend on $\delta$, and the dependence on the fixed chain length $N$ is left implicit; we will write $n_{\mathrm{EB}}(N,\delta)$ when comparing different chain lengths. The PPT criterion thus complements the capacity bounds by identifying the exact parameter region in which the effective channel destroys entanglement with any reference system. Through its dependence on $\gamma_n$, $p_n$, and $\lambda_n$, it captures the combined influence of population-transfer loss, excitation imbalance, and memory-induced dephasing. In Sec. IV, we apply these tools to quantify how chain length, repeated use, and readout timing errors constrain quantum communication and entanglement distribution.

\section{Results}

We first establish timing-error thresholds for zero unassisted quantum capacity and positive-capacity certification, including the maximal tolerance certified on the second use across chains with internal memory. These thresholds guide the subsequent analysis of representative timing-error regimes, where we examine the joint influence of chain length and channel reuse on the capacity window. Finally, we apply the analytical PPT criterion to identify entanglement-breaking regimes and determine the corresponding critical reuse numbers.

\subsection{Zero-capacity and certification thresholds}

We first establish timing-error thresholds that provide benchmarks for the
subsequent analysis of channel reuse. We consider the engineered
perfect-state-transfer chain studied in Ref.~\cite{Zammit:2025orr}, with couplings
$J_j \propto \sqrt{j(N-j)}$, and assume the same readout duration at every
use. Writing $t=\tau(1\pm\delta)$, the damping parameter becomes
\begin{equation}
\gamma_n^{(N)}(\delta)
=\gamma^{(N)}(\delta)
=1-\cos^{2(N-1)}\left(\frac{\pi\delta}{2}\right),
\end{equation}
which depends on chain length and timing error but is independent of the
channel-use number.

The damping-induced antidegradability criterion discussed in Ref.~\cite{Zammit:2025orr}
guarantees zero unassisted quantum capacity whenever
$\gamma^{(N)}(\delta)\ge 1/2$. Solving this inequality gives the
chain-length-dependent threshold
\begin{equation}
\delta_*(N)
=\frac{2}{\pi}
\arccos\left(2^{-1/[2(N-1)]}\right),
\end{equation}
such that
\begin{equation}
\delta\ge\delta_*(N)
\quad\Longrightarrow\quad
Q\left(\Phi_n^{(N)}(\delta)\right)=0
\qquad\text{for every }n.
\end{equation}
The threshold decreases with chain length. Among chains with internal
memory, $N\ge 3$, its largest value is therefore attained by the three-site
chain:
\begin{equation}
\delta_*(3)
=\frac{2}{\pi}\arccos\left(2^{-1/4}\right)
\simeq 36.4057\%.
\end{equation}
Consequently, relative timing errors of $36.41\%$ or higher guarantee zero
unassisted quantum capacity for every chain with internal memory,
independently of the number of uses.

\begin{figure*}[t!]
    \centering
    \begin{subfigure}[b]{0.32\textwidth}
        \centering
        \includegraphics[width=\textwidth]{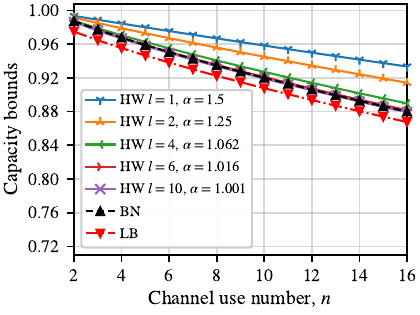}
        \caption{$\delta = 1\%$}
        \label{fig:bounds_01}
    \end{subfigure}
    \hfill
    \begin{subfigure}[b]{0.32\textwidth}
        \centering
        \includegraphics[width=\textwidth]{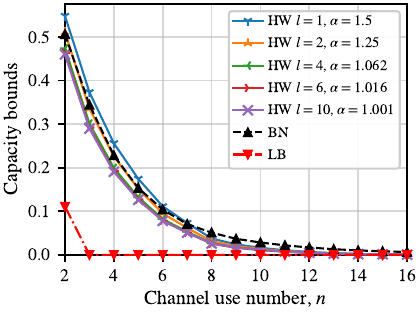}
        \caption{$\delta = 10\%$}
        \label{fig:bounds_10}
    \end{subfigure}
    \hfill
    \begin{subfigure}[b]{0.32\textwidth}
        \centering
        \includegraphics[width=\textwidth]{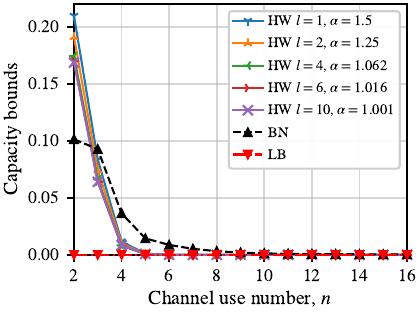}
        \caption{$\delta = 16\%$}
        \label{fig:bounds_16}
    \end{subfigure}
    \caption{Analytical lower bound $\mathrm{LB}(n)$, bottleneck bound
      $\mathrm{BN}(n)$, and Holevo--Werner upper bounds $\mathrm{HW}(n, l)$ as
      functions of the channel-use number $n$ for chain length $N = 10$ at
      relative readout timing errors $\delta \in \{1\%, 10\%, 16\%\}$.}
    \label{fig:bounds_vs_n}
\end{figure*}

To determine the complementary regime in which positive capacity remains
certified, we next examine the positivity of our lower bound. For each
chain length and use number, we define the certification threshold
\begin{equation}
\delta_{\mathrm{LB}}^{(N)}(n)
:=\sup\left\{
\delta\in[0,1]:
\mathrm{LB}^{(N)}(n,\delta)>0
\right\}.
\end{equation}

On the second use, the maximal timing-error tolerance certified by our
lower bound across all chains with internal memory is attained by the
three-site chain:
\begin{equation}
\delta_{\mathrm{LB}}^{(N)}(2)
\leq
\delta_{\mathrm{LB}}^{(3)}(2)
\simeq 22.624\%,
\qquad N\ge 3.
\end{equation}
Consequently, the maximal relative timing-error tolerance certified by
our lower bound on the second use lies just below $22.63\%$.

The three-site threshold is obtained by numerical maximization of the
analytical lower bound. Its optimality across chain lengths follows from
an analytical comparison: we construct an upper bound on
$\mathrm{LB}^{(N)}(2,\delta)$ that is nonincreasing with $N$ and, for every
$N\ge 4$, vanishes at smaller timing errors than the three-site
certification threshold. The detailed construction, comparison, and
numerical evaluation are provided in Appendix~D.

\subsection{Capacity bounds across chain length and reuse}
Having established the general timing-error thresholds in Sec. IV A, we now examine how the capacity bounds vary jointly with chain length, reuse number, and readout timing error. We evaluate the analytical lower bound $\mathrm{LB}^{(N)}(n)$, the bottleneck upper bound $\mathrm{BN}^{(N)}(n)$, and the Holevo--Werner semidefinite-programming upper bound $\mathrm{HW}^{(N)}(n,l)$ for the effective channels $\Phi_n^{(N)}$ generated when the same spin chain is reused without reinitialization. As established in Sec. III, these quantities characterize independent realizations of each fixed effective map $\Phi_n^{(N)}$, rather than collective coding across the correlated sequence of physical channel uses. Under the equal-readout-time protocol, the damping parameter $\gamma^{(N)}(\delta)$ depends on the chain length and timing error but is independent of the reuse number, whereas $p_n^{(N)}$ and $\lambda_n^{(N)}$ encode the excitation imbalance and dephasing accumulated from previous transfers. Together, these parameters determine the generalized-amplitude-damping and dephasing contributions and hence the capacity window in Eq.~(\ref{capwind}). We study chains with internal memory of length $N=3,\ldots,10$ over uses $n=2,\ldots,16$ at representative timing errors. As discussed in Sec.~II B, the case $N=2$ is excluded because no internal spin remains between successive transfers and the corresponding channel therefore has no reuse-induced memory.  Our analysis distinguishes regions of certified positive capacity, unresolved capacity, and certified zero capacity.

\begin{figure*}[t!]
    \centering
    \begin{subfigure}[b]{0.32\textwidth}
        \centering
        \includegraphics[width=\textwidth]{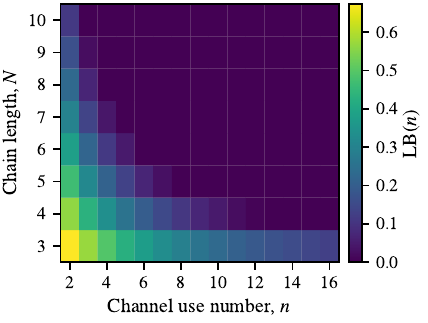}
        \caption{$\delta = 10\%$}
        \label{fig:heatmap_lb_10}
    \end{subfigure}
    \hfill
    \begin{subfigure}[b]{0.32\textwidth}
        \centering
        \includegraphics[width=\textwidth]{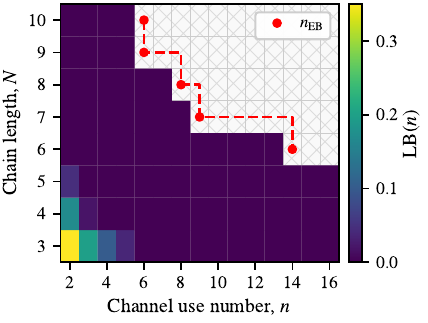}
        \caption{$\delta = 16\%$}
        \label{fig:heatmap_lb_16}
    \end{subfigure}
    \hfill
    \begin{subfigure}[b]{0.32\textwidth}
        \centering
        \includegraphics[width=\textwidth]{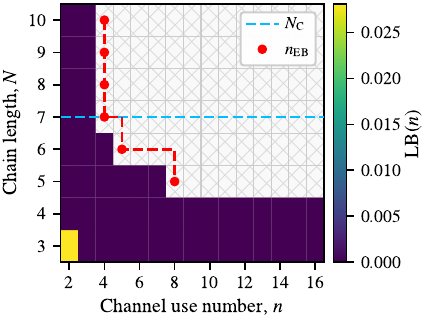}
        \caption{$\delta = 22\%$}
        \label{fig:heatmap_lb_22}
    \end{subfigure}
    \caption{Analytical lower bound $\mathrm{LB}(n)$ across channel use number
      $n$ and chain length $N$ for relative readout timing errors
      $\delta \in \{10\%, 16\%, 22\%\}$. The red line indicates the
      entanglement-breaking threshold $n_{\mathrm{EB}}$ (the hatched region
      corresponds to $n \geq n_{\mathrm{EB}}$), and the blue line marks the critical chain length $N_C$, at or above which $\gamma^{(N)}(\delta)\geq1/2$.}
    \label{fig:heatmap_LB}
\end{figure*}

\begin{figure*}
    \centering
    \begin{subfigure}[b]{0.32\textwidth}
        \centering
        \includegraphics[width=\textwidth]{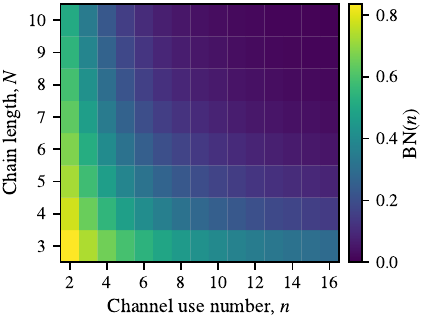}
        \caption{$\delta = 10\%$}
        \label{fig:heatmap_bn_10}
    \end{subfigure}
    \hfill
    \begin{subfigure}[b]{0.32\textwidth}
        \centering
        \includegraphics[width=\textwidth]{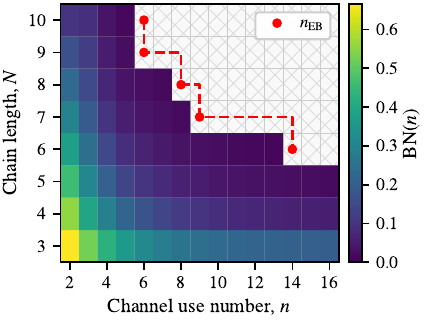}
        \caption{$\delta = 16\%$}
        \label{fig:heatmap_bn_16}
    \end{subfigure}
    \hfill
    \begin{subfigure}[b]{0.32\textwidth}
        \centering
        \includegraphics[width=\textwidth]{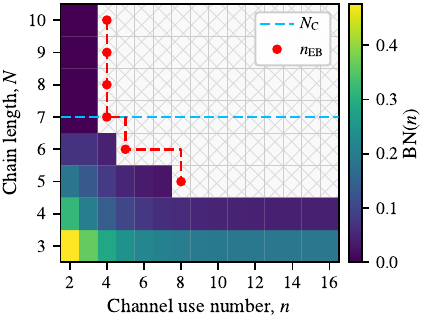}
        \caption{$\delta = 22\%$}
        \label{fig:heatmap_bn_22}
    \end{subfigure}
    \caption{Bottleneck upper bound $\mathrm{BN}(n)$ across channel-use number $n$ and chain length $N$ for relative readout timing
      errors $\delta\in\{10\%,16\%,22\%\}$. The red staircase marks the entanglement-breaking boundary $n_{\mathrm{EB}}(N,\delta)$, with $n\geq n_{\mathrm{EB}}$ hatched. The blue dashed line marks the critical chain length $N_C$, at or above which $\gamma^{(N)}(\delta)\geq1/2$ and $\mathrm{BN}(n)=0$ for every $n$.}
    \label{fig:heatmap_BN}
\end{figure*}

\begin{figure*}
    \centering
    \begin{subfigure}[b]{0.32\textwidth}
        \centering
        \includegraphics[width=\textwidth]{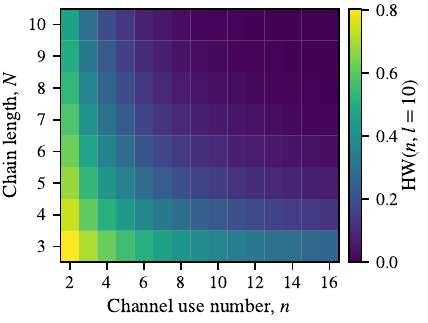}
        \caption{$\delta = 10\%$}
        \label{fig:heatmap_hw_10}
    \end{subfigure}
    \hfill
    \begin{subfigure}[b]{0.32\textwidth}
        \centering
        \includegraphics[width=\textwidth]{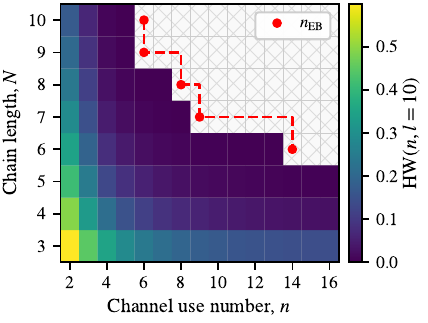}
        \caption{$\delta = 16\%$}
        \label{fig:heatmap_hw_16}
    \end{subfigure}
    \hfill
    \begin{subfigure}[b]{0.32\textwidth}
        \centering
        \includegraphics[width=\textwidth]{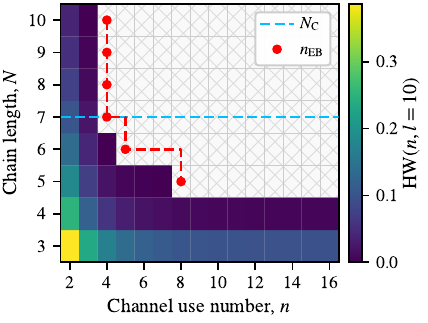}
        \caption{$\delta = 22\%$}
        \label{fig:heatmap_hw_22}
    \end{subfigure}
    \caption{Holevo--Werner SDP upper bound $\mathrm{HW}(n, l=10)$ across
      channel-use number $n$ and chain length $N$ for relative readout timing
      errors $\delta \in \{10\%, 16\%, 22\% \}$. The red staircase marks the entanglement-breaking boundary $n_{\mathrm{EB}}(N,\delta)$, while the blue dashed line marks the critical length $N_C$, at or above which $\gamma^{(N)}(\delta)\geq1/2$. Hatched cells satisfy $n\geq n_{\mathrm{EB}}$ and therefore belong to a zero-capacity region.}
    \label{fig:heatmap_HW}
\end{figure*}

The resulting bounds reveal two complementary mechanisms by which quantum transmission deteriorates. At fixed $N$ and $\delta$, increasing the reuse number modifies $p_n^{(N)}$ and $\lambda_n^{(N)}$ through the residual excitations left by previous transfers, generally shifting the lower and upper bounds toward zero and reducing the region in which positive capacity is certified. At fixed $n$ and $\delta$, increasing the chain length enhances the population-transfer loss through $\gamma^{(N)}(\delta)$. Once $N$ reaches the critical length $N_C(\delta)$, the condition $\gamma^{(N)}(\delta)\geq 1/2$ causes the bottleneck bound to vanish for every $n$. Repeated-use degradation is therefore governed jointly by memory accumulation and length-dependent damping, rather than by the reuse number alone.

Figure~\ref{fig:bounds_vs_n} presents representative one-dimensional cuts at fixed chain length $N=10$ for $\delta=1\%$, $10\%$, and $16\%$. These values respectively probe near-ideal transmission, pronounced degradation under reuse, and strong suppression without positive-capacity certification by the lower bound. Importantly, all three lie below the $N=10$ damping threshold $\delta_*(10)\simeq17.55\%$, so that $\gamma^{(10)}(\delta)<1/2$ and the bottleneck bound is not forced to vanish for every use by the damping contribution alone. The comparison therefore isolates how memory accumulation progressively modifies the capacity bounds as $n$ increases. For $\delta=1\%$, all bounds remain positive throughout the displayed range and decrease only gradually with $n$. At $\delta=10\%$, the degradation is substantially faster: the lower bound certifies positive capacity only for the second use, whereas the bottleneck and Holevo--Werner bounds decay toward zero with increasing $n$. At $\delta=16\%$, the lower bound vanishes throughout the displayed range, while the upper bounds remain appreciable only during the earliest uses. A vanishing lower bound does not establish zero capacity; it instead identifies a region in which the present achievable rate no longer certifies positive capacity. Across all three panels, increasing the hierarchy level $l$ systematically tightens the Holevo--Werner converse, with $l=10$ providing the strongest level evaluated here. We therefore use $\mathrm{HW}(n,10)$ in the two-dimensional analysis below.

The heat maps in Figs.~\ref{fig:heatmap_LB}--\ref{fig:heatmap_HW} extend these fixed-length cuts to the full $(N,n)$ landscape. Figure~\ref{fig:heatmap_LB} shows that, at $\delta=10\%$, the region in which the lower bound certifies positive capacity has a wedge-like structure: short chains retain a nonzero achievable rate over many uses, whereas increasing $N$ rapidly restricts certification to the earliest uses. At $\delta=16\%$, this region contracts sharply toward small $N$ and small $n$, and at $\delta=22\%$ positive capacity is certified only at $(N,n)=(3,2)$, consistently with the maximal second-use threshold $\delta_{\mathrm{LB}}^{(3)}(2)\simeq22.624\%$. The bottleneck and Holevo--Werner landscapes in Figs.~\ref{fig:heatmap_BN} and \ref{fig:heatmap_HW} exhibit the same overall suppression with increasing chain length, reuse number, and timing error, but remain nonzero over a broader domain than the lower bound. The computed Holevo--Werner bound lies systematically below the bottleneck bound, providing a tighter converse where damping and dephasing act jointly. At $\delta=22\%$, the critical length $N_C=7$ produces an $n$-independent zero-capacity sector for $N\geq7$, while the staircase-shaped entanglement-breaking boundary extends the certified zero-capacity region toward shorter chains as the reuse number increases. Points for which the lower bound vanishes while both upper bounds remain positive and the PPT condition is not satisfied constitute an unresolved region, rather than evidence of zero capacity.

\begin{figure*}
    \centering
    \begin{subfigure}[b]{0.32\textwidth}
        \centering
        \includegraphics[width=\textwidth]{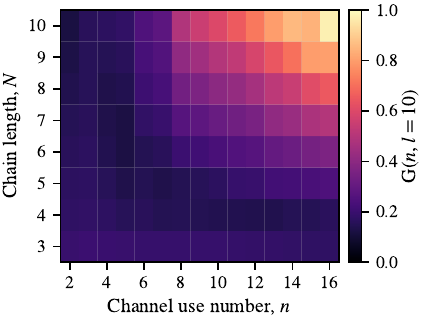}
        \caption{$\delta = 10\%$}
        \label{fig:heatmap_g_10}
    \end{subfigure}
    \hfill
    \begin{subfigure}[b]{0.32\textwidth}
        \centering
        \includegraphics[width=\textwidth]{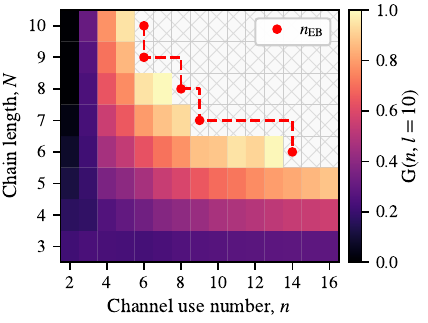}
        \caption{$\delta = 16\%$}
        \label{fig:heatmap_g_16}
    \end{subfigure}
    \hfill
    \begin{subfigure}[b]{0.32\textwidth}
        \centering
        \includegraphics[width=\textwidth]{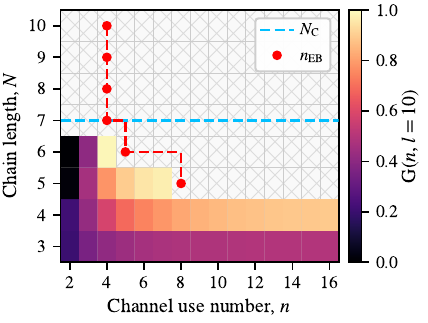}
        \caption{$\delta = 22\%$}
        \label{fig:heatmap_g_22}
    \end{subfigure}
    \caption{Relative gap-reduction factor $\mathrm{G}(n,l=10)$ across channel-use number $n$ and chain length $N$ for representative relative readout timing
      errors $\delta \in \{10\%, 16\%, 22\% \}$. Hatched cells are excluded from the comparison: for $n\geq n_{\mathrm{EB}}$ (red boundary), the channel is entanglement breaking, whereas for $N\geq N_C$ (blue boundary), $\mathrm{BN}=\mathrm{LB}=0$ and $\mathrm{G}$ is undefined.}
    \label{fig:heatmap_G}
\end{figure*}

Additionally, to quantify how much the Holevo--Werner SDP upper bound tightens the analytical bottleneck bound across chain length and reuse number, we introduce the relative gap reduction factor $\mathrm{G}\left(n,l\right)$. For each fixed $N$ and $\delta$, this quantity measures the fraction of the original analytical capacity window, $\mathrm{BN}(n)-\mathrm{LB}(n)$, that is closed by replacing $\mathrm{BN}(n)$ with $\mathrm{HW}\left(n,l\right)$:
\begin{equation}
\mathrm{G}\left(n,l\right)=\frac{\mathrm{BN}(n)-\mathrm{HW}\left(n,l\right)}{\mathrm{BN}(n)-\mathrm{LB}(n)}
\end{equation}
Thus, $\mathrm{G}\left(n,l\right)=0$ corresponds to no improvement over the analytical upper bound, whereas $\mathrm{G}\left(n,l\right)=1$ indicates that the SDP upper bound reaches the lower bound and closes the entire original capacity window. Intermediate values quantify the fraction of that window closed by the Holevo--Werner converse. Rather than being controlled by the timing error alone, the gap reduction depends jointly on $\delta$, $N$, and $n$. It becomes most pronounced near the length- and reuse-dependent zero-capacity boundaries, where the combined action of damping and dephasing is not fully captured by the analytical bottleneck bound.

Figure~\ref{fig:heatmap_G} shows that the relative tightening provided by the Holevo--Werner bound depends jointly on the chain length $N$, the channel-use number $n$, and the relative readout timing error $\delta$. The values $\delta=10\%$, $16\%$, and $22\%$ are representative examples; our numerical evaluation over the full admissible timing-error range for which the normalized gap reduction is defined exhibits the same qualitative behavior. At $\delta=10\%$, the gap reduction is modest for short chains and early uses but increases toward the large-$N$, late-use region, where the SDP bound closes most of the analytical capacity window. At $\delta=16\%$, a region of large gap reduction develops immediately below the entanglement-breaking boundary, indicating that the Holevo--Werner bound becomes particularly informative as the effective channel approaches this transition. At $\delta=22\%$, this region shifts toward shorter chains and earlier uses, while for $N\geq N_C=7$ the bottleneck bound already vanishes and the gap comparison is no longer informative. The improvement provided by the SDP converse is therefore not governed by a single timing-error crossover; instead, it becomes strongest near the length- and reuse-dependent zero-capacity boundaries, where the combined action of damping and dephasing is not fully captured by the separate-channel bottleneck bound.

Taken together, these landscapes show a progressive contraction of the region in which nonzero quantum capacity is certified. The bounds nevertheless leave an intermediate region unresolved: a vanishing lower bound is inconclusive unless one of the upper bounds also vanishes. The PPT criterion provides a complementary exact characterization of the parameter region in which the effective channel is entanglement breaking and therefore has zero quantum capacity. In the following subsection, we analyze the corresponding critical reuse boundary $n_{\mathrm{EB}}$ and its dependence on chain length and readout timing error.

\subsection{Entanglement-breaking boundary}

\begin{figure}
    \centering
    \includegraphics[width=\linewidth]{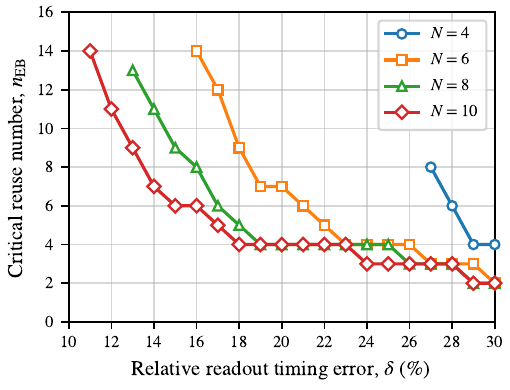}
    \caption{Critical reuse number $n_{\mathrm{EB}}$ as a function of the
      relative readout timing error $\delta$ for chain lengths
      $N \in \{4, 6, 8, 10\}$. Points with $n_{\mathrm{EB}} > 16$ are omitted.
      Lines are guides to the eye.}
    \label{fig:critical_lengths}
\end{figure}

\begin{figure}
    \centering
    \includegraphics[width=\linewidth]{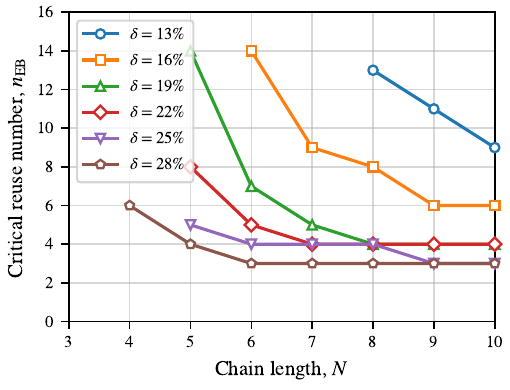}
    \caption{Critical reuse number $n_{\mathrm{EB}}$ as a function of chain
      length $N$ for relative readout timing errors
      $\delta \in \{13\%, 16\%, 19\%, 22\%, 25\%, 28\%\}$. Points with
      $n_{\mathrm{EB}} > 16$ are omitted. Lines are guides to the eye.}
    \label{fig:critical_vs_N}
\end{figure}
Figure~\ref{fig:critical_lengths} shows the critical reuse number $n_{\mathrm{EB}}(N,\delta)$ as a function of the relative readout timing error $\delta$ for representative chain lengths $N=4$, $6$, $8$, and $10$. For every chain length, increasing $\delta$ moves the entanglement-breaking transition toward earlier uses, but the onset depends strongly on $N$. For $N=10$, no transition is observed within $n\leq16$ for errors up to $10\%$; the transition occurs at $n_{\mathrm{EB}}=14$, $9$, $6$, $4$, and $3$ for $\delta=11\%$, $13\%$, $15\%$, $20\%$, and $28\%$, respectively, and reaches the second use at $\delta=29\%$. Shorter chains are substantially more robust: the first transitions visible within the investigated range occur at $\delta=13\%$ for $N=8$, $\delta=16\%$ for $N=6$, and only at $\delta=27\%$ for $N=4$. Points for which $n_{\mathrm{EB}}>16$ are omitted.

The complementary representation in Fig.~\ref{fig:critical_vs_N} shows $n_{\mathrm{EB}}(N,\delta)$ as a function of chain length for the representative fixed timing errors $\delta\in\{13\%,16\%,19\%,22\%,25\%,28\%\}$. These values are selected to illustrate the evolution of the boundary; our numerical evaluation over the full timing-error range considered exhibits the same qualitative behavior, with $n_{\mathrm{EB}}$ decreasing or remaining constant as $N$ increases. At $\delta=13\%$, a transition within the investigated range occurs only for $N\geq8$, with $n_{\mathrm{EB}}$ decreasing from $13$ at $N=8$ to $9$ at $N=10$. At $\delta=16\%$, the transition appears at $n_{\mathrm{EB}}=14$ for $N=6$ and advances to $n_{\mathrm{EB}}=6$ for $N=10$. At $\delta=28\%$, it occurs by the sixth use for $N=4$, by the fourth use for $N=5$, and by the third use for every $N\geq6$. These results show that the entanglement-breaking boundary is governed jointly by timing error, chain length, and reuse number. Increasing the chain length enhances the damping associated with imperfect readout, while repeated use modifies the excitation imbalance and dephasing through the memory retained from previous transfers. The combined action of these effects moves the entanglement-breaking transition toward earlier uses as either $N$ or $\delta$ increases.

\section{Discussion and conclusions}

We characterized quantum-information transmission through a spin-chain channel reused without complete reinitialization, resolving its dependence on chain length $N$, reuse number $n$, and relative readout timing error $\delta$. Each effective channel can be represented as dephasing followed by generalized amplitude damping. Under the equal-readout-time protocol, $N$ and $\delta$ fix the damping parameter, while the excitation imbalance and dephasing parameters retain the history of previous transfers. The resulting degradation therefore arises from the joint action of length-dependent population-transfer loss and reuse-induced memory accumulation.

The capacity bounds reveal information that cannot be inferred from state-transfer fidelity alone. Reduced fidelity does not necessarily imply vanishing quantum capacity, while a channel with nonzero transfer fidelity may nevertheless be unable to support asymptotically reliable quantum communication \cite{Barnum1998,Lloyd1997,Khatri:2020dfc}. The combined lower- and upper-bound landscapes distinguish regions of certified positive capacity, unresolved capacity, and certified zero capacity across $N$, $n$, and $\delta$. For short chains, early uses, and small timing errors, the analytical bottleneck bound remains comparatively informative. As the chain length, reuse number, or timing error increases, the Holevo--Werner converse closes a progressively larger fraction of the analytical capacity window, becoming most advantageous near the zero-capacity boundaries. This behavior shows that the combined action of damping and dephasing, which the separate-channel bottleneck bound does not fully capture, is essential in regimes most strongly limited by accumulated memory.

 Moreover, because each effective channel's Choi state is a two-qubit state, positivity under partial transposition (PPT) is equivalent to separability, thus certifying that the channel is entanglement-breaking and has zero quantum capacity. Mapping the corresponding critical reuse number $n_{\mathrm{EB}}$ over chain length and timing error reveals a systematic trend: $n_{\mathrm{EB}}$ decreases as either $N$ or $\delta$ increases. Short chains and small timing errors can therefore remain outside the entanglement-breaking region over many reuses, whereas longer chains or larger errors drive the transition toward progressively earlier uses. The transition is thus governed by a joint length--error--reuse relation, reflecting the cumulative interplay between length-dependent damping and memory retained from previous transfers.

 Additionally, our analysis shows that each timing error sets a critical chain length $N_C(\delta)$ beyond which damping alone forces the unassisted quantum capacity to vanish for every reuse, while for shorter chains accumulated memory can still drive the effective channel toward entanglement breaking as $n$ increases. Two global benchmarks delimit this behavior: the maximal second-use timing-error tolerance certified by the lower bound is just below $22.63\%$ and is attained by the three-site chain, whereas errors of $36.41\%$ or greater guarantee zero unassisted quantum capacity for every chain with internal memory and every reuse number. Between these benchmarks, the outcome depends jointly on $N$, $n$, and $\delta$. Timing errors and incomplete reinitialization therefore act cooperatively, so imperfections that are moderate in a single transfer can accumulate under channel reuse and eventually prevent entanglement distribution. Furthermore, whenever the Holevo--Werner bound vanishes, it also rules out a nonzero asymptotic rate under PPT-assisted, and hence LOCC-assisted, entanglement distillation \cite{Horodecki2009Review,HolevoWerner2001,FangFawzi2021}. Therefore, post-processing independent realizations of the fixed effective map $\Phi_n$ cannot overcome such a capacity collapse.

Our conclusions apply when the internal channel memory and
the timing-error realization are inaccessible to the encoder
and decoder. They do not exclude recovery through direct
control of the memory or active channel reinitialization
\cite{carusoQuantumChannelsMemory2014b,navascuesResettingUncontrolledQuantum2018,Giovannetti_2005,PhysRevA.69.012306}. Such resources introduce additional temporal and control overheads, producing a trade-off between rapid channel reuse and mitigation of memory-induced degradation that is relevant to quantum repeaters and modular quantum links \cite{Azuma2023,Awschalom2021}.

Recent experiments illustrate the practical relevance of this
regime, including heralded entanglement distribution between
solid-state nodes over deployed metropolitan fiber
\cite{Stolk2024}, memory--memory entanglement between remote
quantum-network nodes \cite{Liu2024}, and device-independent
quantum key distribution between distant matter-qubit systems
\cite{Zhang2022}. In such platforms, entanglement must be generated not only
with high fidelity in a single attempt, but repeatedly and at
useful rates under finite memory lifetimes, synchronization
requirements, and reset overheads
\cite{Sangouard2011, Magnard2018,
Stolk2024Synchronization}.

 This work provides an operational benchmark for the considered setting by relating readout timing errors, chain length, and residual chain excitations to the progressive contraction of the quantum-capacity window and, ultimately, to the onset of entanglement breaking. Our results thereby connect microscopic memory accumulation to a link-level constraint on reliable entanglement distribution and provide a criterion for assessing when active reset, improved synchronization, or strategies for mitigating memory-induced degradation become necessary for scalable quantum networks.

\begin{acknowledgments}

HZ, GV, TJGA, and JB acknowledge funding from Grant Agreement No.~DTP-2024-13, Project ATTESTER financed by the Xjenza Malta, and the Malta Digital Innovation Authority, through the `R\&I Thematic Programmes: Digital Technologies Programme'.  RS and TS acknowledge funding from the European Union's Horizon Europe research and innovation programme under the
project ``Quantum Secure Networks Partnership''
(QSNP, grant agreement No 101114043). HZ also acknowledges funding by the Tertiary Education Scholarships Scheme (TESS).
\end{acknowledgments}

\section*{Data Availability}

The numerical data supporting the findings of this article have not been deposited in a public repository because they can be regenerated from the equations and numerical procedures described in the article. The numerical data and custom code used to generate the figures are available from the corresponding author upon reasonable request. No experimental or externally sourced data were used.

\appendix

\section{Degradability properties of the effective channel}
This appendix determines all cases in which the effective dynamical map is degradable. A channel $\mathcal E$ is degradable if there exists a channel $\mathcal P$ such that $\mathcal E^c=\mathcal P\circ\mathcal E$, where $\mathcal E^c$ is a complementary channel. For a degradable channel, the coherent information is additive, and the quantum capacity therefore reduces to eq. (\ref{chandeg}).

Now, for the quantum
qubit channel, it is known to be degradable if and only if the
Kraus rank is at most 2 and the Kraus operators are -up to unitaries-
of the form \cite{Wolf2007}:
\begin{equation}
\tilde{K}_{1}=\left(\begin{array}{cc}
\cos\left(\alpha\right) & 0\\
0 & \cos\left(\beta\right)
\end{array}\right),\quad\tilde{K}_{2}=\left(\begin{array}{cc}
0 & \sin\left(\beta\right)\\
\sin\left(\alpha\right) & 0
\end{array}\right).\label{eq:Iffcond}
\end{equation}
For $\alpha=\beta$ the above represents a dephasing channel and for
$\beta=0$ an amplitude damping channel. 

Employing the decomposition of the $n$-use channel into a general
 amplitude damping $ \mathcal{A}_{\gamma_{n},p_{n}}$ and dephasing $\mathcal{D}_{\lambda_{n}} $ channel:

\begin{equation}
\mathcal{E}_{\Phi_n}=\mathcal{A}_{\gamma_{n},p_{n}}\circ\mathcal{D}_{\lambda_{n}}
\end{equation}
\begin{widetext}
We can find the Jamiolkowski state of $\mathcal{E}^{\left(n\right)}$
is: 
\begin{equation}
J\left(\mathcal{E}_{\Phi_n}\right)=\frac{1}{2}\left(\begin{array}{cccc}
1-\gamma_{n}(1-p_{n}) & 0 & 0 & \sqrt{(1-\gamma_{n})(1-\lambda_{n})}\\
0 & \gamma_{n}(1-p_{n}) & 0 & 0\\
0 & 0 & \gamma_{n}p_{n} & 0\\
\sqrt{(1-\gamma_{n})(1-\lambda_{n})} & 0 & 0 & 1-\gamma_{n}p_{n}
\end{array}\right)
\end{equation}
\end{widetext}
with the four eigenvalues: 
\begin{eqnarray*}
\mu_{\pm} & = & \frac{1}{2}\left[\!1\!-\!\frac{\gamma_{n}}{2}\!\pm\!\sqrt{\!\left(\frac{\gamma_{n}}{2}\right)^{2}\!\!\left(1-2p_{n}\right)^{2}\!\!+\!(1\!-\!\gamma_{n})(1\!-\!\lambda_{n})}\right]\\
\mu_{1} & = & \frac{1}{2}\gamma_{n}p_{n}\\
\mu_{2} & = & \frac{1}{2}\gamma_{n}(1-p_{n})
\end{eqnarray*}
Then, for $\mathcal{E}_{\Phi_n}$ to satisfy a decomposition in Kraus operators as in
(\ref{eq:Iffcond}), its Jamiolkowski state $J\left(\mathcal{E}_{\Phi_n}\right)$
must have at most two eigenvalues different from zero. By exhaustively
analyzing the four eigenvalues $\mu_{\pm},\mu_{1},\mu_{2}$ we find
that this happens only when either $\gamma_{n}=0$ (i.e. no-damping)
or $\lambda_{n}=0$ and $p_{n}\in\left\{ 0,1\right\} $ (i.e. no dephasing
and extremal amplitude damping). Later, in the first case the Kraus
operators take the form (\ref{eq:Iffcond}) for any value of $\lambda_{n}\in\left[0,1\right]$
and $p_{n}$ is irrelevant, while in the second case the condition
(\ref{eq:Iffcond}) is satisfied for $\gamma_{n}\leq1/2$, in both
cases $p=0$ or $p=1$, where it reduces to the limit case of not-general
amplitude damping channels \cite{Khatri2020}. 

Summarizing the $n$-use use channel $\mathcal{E}_{\Phi_n}$
is degradable if and only if it is a usual amplitude damping $\mathcal{A}_{\gamma_n,0}$
or $\text{\ensuremath{\mathcal{A}_{\gamma_n,1}}}$ with $\gamma_n\leq1/2$
or a purely dephasing channel $\mathcal{D}_{\lambda_n}$, but not both. 

Furthermore, equation (84) of reference \cite{Khatri2020} gives the convex decomposition of the GAD channel into the two extremal amplitude-damping channels:
\begin{equation}
    \mathcal{A}_{\gamma_{n},p_{n}} = p_{n}\mathcal{A}_{\gamma_{n},0} + \left(1-p_{n}\right)\mathcal{A}_{\gamma_{n},1}~,
\end{equation}
Since coherent information is convex in the channel for a fixed input, and both extremal channels are degradable for $\gamma\leq 1/2$, this yields

\begin{equation}
    Q(\mathcal{A}_{\gamma_{n},p_{n}}) \leq p_{n}Q(\mathcal{A}_{\gamma_{n},0}) + \left(1-p_{n}\right)Q(\mathcal{A}_{\gamma_{n},1})~.
\end{equation}
Because $\mathcal A_{\gamma,0}$ and $\mathcal A_{\gamma,1}$ are unitarily equivalent, their capacities coincide. Therefore,
\begin{equation}
    Q(\mathcal{A}_{\gamma_{n},p_{n}}) \leq Q(\mathcal{A}_{\gamma_{n},0}) =I_{c} \left[\mathcal{A}_{\gamma_{n},0}\right],
\end{equation}
holds over the full range $\gamma_n\in[0,1]$: below the antidegradability threshold it follows from the convex decomposition into degradable extremal channels, while at and above the threshold both sides vanish because for $\gamma_n\geq1/2$, the generic GAD channel is antidegradable for every $p_n$ \cite{Khatri2020}. Finally, due to the degradability of the dephasing channel \cite{Devetak2005},
\begin{equation}
Q(\mathcal{D}_{\lambda_{n}}) \equiv I_{c}\left[\mathcal{D}_{\lambda_{n}}\right]~. 
\end{equation}

\section{Derivation of the analytical lower bound}
In this appendix we include the complete proof of the analytical lower bound  (\ref{eq:lowerboundZ}), by maximizing the coherent information at the subset of diagonal states, i.e. computing $S\left(\mathcal{A}_{\gamma,p}\!\circ\!\mathcal{D}_{\lambda}\left(\rho_{z}\right)\right)$
and $S\left(\left[\mathcal{A}_{\gamma,p}\!\circ\!\mathcal{D}_{\lambda}\right]^{c}\left(\rho_{z}\right)\right)$
with $\rho_{z}$ a diagonal state of the form:
\begin{equation}
\rho_{z}=\frac{1}{2}\left(I+z\sigma_{z}\right)=\left(\begin{array}{cc}
a_{+} & 0\\
0 & a_{-}
\end{array}\right),\label{eq:diagstate}
\end{equation}
and then maximize over $z\in\left[-1,1\right]$. 

Computing the first
term we have:
\begin{eqnarray*}
S\left(\mathcal{A}_{\gamma,p}\!\circ\!\mathcal{D}_{\lambda}\left(\rho_{z}\right)\right) & \overset{(a)}{=} & S\left(\mathcal{A}_{\gamma,p}\left(\rho_{z}\right)\right)\\
 & \overset{(b)}{=} & S\left(\rho_{z^{\prime}}\right)\\
 & = & h\left(\frac{1+z^{\prime}}{2}\right)
\end{eqnarray*}
where in $(a)$ we use the fact that diagonal states are invariant
under dephasing channels $\mathcal{D}_{\lambda}$, and in $(b)$ that
general amplitude channels $\mathcal{A}_{\gamma,p}$ only apply an
affine transformation $z^{\prime}=(1-\gamma)z+\gamma(2p-1)$ to the
$z$ parameter of diagonal states $\rho_{z}$. Consequently:
\begin{equation}
S\left(\mathcal{A}_{\gamma,p}\!\circ\!\mathcal{D}_{\lambda}\left(\rho_{z}\right)\right)=h\!\left(\frac{1+(1-\gamma)z+\gamma(2p-1)}{2}\right).
\end{equation}
Now, to calculate the term $S\left(\left[\mathcal{A}_{\gamma,p}\!\circ\!\mathcal{D}_{\lambda}\right]^{c}\left(\rho_{z}\right)\right)$
we need to employ equation (\ref{eq:complchan}), and thus we first
recall the Kraus operators of $\mathcal{A}_{\gamma,p}$ and $\mathcal{D}_{\lambda}$. 

We start with the dephasing channel $\mathcal{D}_{\lambda}$ employing
the following Kraus operator representation:
\begin{equation}
K_{+}=\eta_{+}I,\quad K_{-}=\eta_{-}\sigma_{z},
\end{equation}
where:
\begin{equation}
\eta_{\pm}=\frac{1\pm\sqrt{1-\lambda}}{2}
\end{equation}
while for the general amplitude damping channel $\mathcal{A}_{\gamma,p}$
we use:
\begin{eqnarray}
E_{0}=\sqrt{p}\left(\begin{array}{cc}
1 & 0\\
0 & \mu
\end{array}\right), &  & E_{1}=\sqrt{p}\left(\begin{array}{cc}
0 & \nu\\
0 & 0
\end{array}\right),\nonumber \\
E_{2}=\sqrt{1-p}\left(\begin{array}{cc}
\mu & 0\\
0 & 1
\end{array}\right), &  & E_{3}=\sqrt{1-p}\left(\begin{array}{cc}
0 & 0\\
\nu & 0
\end{array}\right),
\end{eqnarray}
where:
\begin{equation}
\mu:=\sqrt{1-\gamma},\quad\nu:=\sqrt{\gamma},
\end{equation}
Then, we must find the Kraus operators $F_{ij}:=E_{i}K_{j}$ for the
composite channel $\mathcal{A}_{\gamma,p}\circ\mathcal{D}_{\lambda}$,
which they split into the two groups:
\begin{equation}
F_{i+}=E_{i}K_{+}=\eta_{+}E_{i},\quad F_{i-}=E_{i}K_{-}=\eta_{-}E_{i}\sigma_{z},
\end{equation}
Let us compute them. From $E_{0}$ we get:
\begin{equation}
F_{0+}=\eta_{+}\sqrt{p}\left(\begin{array}{cc}
1 & 0\\
0 & \mu
\end{array}\right),\quad F_{0-}=\eta_{-}\sqrt{p}\left(\begin{array}{cc}
1 & 0\\
0 & -\mu
\end{array}\right),
\end{equation}
From $E_{1}$:
\begin{equation}
F_{1+}=\eta_{+}\sqrt{p}\left(\begin{array}{cc}
0 & \nu\\
0 & 0
\end{array}\right),\quad F_{1-}=\eta_{-}\sqrt{p}\left(\begin{array}{cc}
0 & -\nu\\
0 & 0
\end{array}\right),
\end{equation}
From $E_{2}$:
\begin{equation}
F_{2+}=\eta_{+}\sqrt{1-p}\left(\begin{array}{cc}
\mu & 0\\
0 & 1
\end{array}\right),\quad F_{2-}=\eta_{-}\sqrt{1-p}\left(\begin{array}{cc}
\mu & 0\\
0 & -1
\end{array}\right),
\end{equation}
and from $E_{3}$
\begin{equation}
F_{3+}=\eta_{+}\sqrt{1-p}\left(\begin{array}{cc}
0 & 0\\
\nu & 0
\end{array}\right),\quad F_{3-}=\eta_{-}\sqrt{1-p}\left(\begin{array}{cc}
0 & 0\\
\nu & 0
\end{array}\right),
\end{equation}
Now, since the input state $\rho_{z}$ is diagonal of the form (\ref{eq:diagstate}),
when replacing in (\ref{eq:complchan}) we have:
\begin{eqnarray*}
\left[\mathcal{A}_{\gamma,p}\!\circ\!\mathcal{D}_{\lambda}\right]^{c}\left(\rho_{z}\right) & = & \sum_{(ij),(rs)}\mathrm{Tr}\left[F_{ij}\rho_{z}F_{rs}^{\dagger}\right]\left|(ij)\right\rangle \!\left\langle (rs)\right|\\
 & = & a_{+}\sum_{(ij),(rs)}\left\langle 0\right|F_{rs}^{\dagger}F_{ij}\left|0\right\rangle \left|(ij)\right\rangle \!\left\langle (rs)\right|\\
 & + & a_{-}\sum_{(ij),(rs)}\left\langle 1\right|F_{rs}^{\dagger}F_{ij}\left|1\right\rangle \left|(ij)\right\rangle \!\left\langle (rs)\right|
\end{eqnarray*}
and now, we continue by analyzing the action of $F_{ij}$ over the
basis $\left\{ \left|0\right\rangle ,\left|1\right\rangle \right\} $.
Let us start listing the action of the composite Kraus operators on
$\left|0\right\rangle $: 
\begin{eqnarray*}
F_{0+}\left|0\right\rangle =\eta_{+}\sqrt{p}\left|0\right\rangle  &  & F_{0-}\left|0\right\rangle =\eta_{-}\sqrt{p}\left|0\right\rangle \\
F_{1+}\left|0\right\rangle =0 &  & F_{1-}\left|0\right\rangle =0\\
F_{2+}\left|0\right\rangle =\eta_{+}\mu\sqrt{1-p}\left|0\right\rangle  &  & F_{2-}\left|0\right\rangle =\eta_{-}\mu\sqrt{1-p}\left|0\right\rangle \\
F_{3+}\left|0\right\rangle =\eta_{+}\nu\sqrt{1-p}\left|1\right\rangle  &  & F_{3-}\left|0\right\rangle =\eta_{-}\nu\sqrt{1-p}\left|1\right\rangle 
\end{eqnarray*}
So, for input $\left|0\right\rangle $, there are two groups, one
that output $\left|0\right\rangle $ and one that output $\left|1\right\rangle $.
Note, these are orthogonal directions, so their contributions decouple.
Let us continue listing the action of the composite Kraus operators
on $\left|1\right\rangle $: 
\begin{eqnarray*}
F_{0+}\left|1\right\rangle =\eta_{+}\mu\sqrt{p}\left|1\right\rangle  &  & F_{0-}\left|1\right\rangle =-\eta_{-}\mu\sqrt{p}\left|1\right\rangle \\
F_{1+}\left|1\right\rangle =\eta_{+}\nu\sqrt{p}\left|0\right\rangle  &  & F_{1-}\left|1\right\rangle =-\eta_{-}\nu\sqrt{p}\left|0\right\rangle \\
F_{2+}\left|1\right\rangle =\eta_{+}\sqrt{1-p}\left|1\right\rangle  &  & F_{2-}\left|1\right\rangle =-\eta_{-}\sqrt{1-p}\left|1\right\rangle \\
F_{3+}\left|1\right\rangle =0 &  & F_{3-}\left|1\right\rangle =0
\end{eqnarray*}
So, for input $\left|1\right\rangle $, there are two groups, one
that output $\left|0\right\rangle $ and one that output $\left|1\right\rangle $.
Again, these are orthogonal directions, so their contributions decouple.
Combining the above observations, we find that the full complementary
state splits into three orthogonal sectors:
\begin{description}
\item [{Coherent}] A coherent sector $\mathbf{C}$ that for input $\left|i\right\rangle $
outputs $\left|i\right\rangle $, generated by the action of the operators
$\left\{ F_{0\pm},F_{2\pm}\right\} $.
\item [{Swap01}] A sector $\mathbf{J}_{0}$ that only comes from input
$\left|0\right\rangle $ and outputs $\left|1\right\rangle $, generated
by the action of the operators $\left\{ F_{3\pm}\right\} $.
\item [{Swap10}] A sector $\mathbf{J}_{1}$ that only comes from input
$\left|1\right\rangle $ and outputs $\left|0\right\rangle $, generated
by the action of the operators $\left\{ F_{1\pm}\right\} $.
\end{description}
So the environment matrix is block diagonal:
\begin{equation}
\left[\mathcal{A}_{\gamma,p}\!\circ\!\mathcal{D}_{\lambda}\right]^{c}\left(\rho_{z}\right)=\mathbf{C}\oplus\mathbf{J}_{0}\oplus\mathbf{J}_{1}
\end{equation}
Now, we start with the block $\mathbf{J}_{0}$ spanned by the basis
$\left|(ij)\right\rangle $ with $(ij)\in(3\pm)$,
\begin{eqnarray*}
\mathbf{J}_{0} & = & a_{+}\sum_{(ij),(rs)\in(3\pm)}\left\langle 0\right|F_{rs}^{\dagger}F_{ij}\left|0\right\rangle \left|(ij)\right\rangle \!\left\langle (rs)\right|\\
 & = & a_{+}\gamma\left(1-p\right)\left(\begin{array}{cc}
\eta_{+}^{2} & \eta_{+}\eta_{-}\\
\eta_{+}\eta_{-} & \eta_{-}^{2}
\end{array}\right)
\end{eqnarray*}
which is rank one with eigenvalues $\left\{ a_{+}\gamma\left(1-p\right),0\right\} $.
Next, we continue with the block $\mathbf{J}_{1}$ spanned by the
basis $\left|(ij)\right\rangle $ with $(ij)\in(1\pm)$,
\begin{eqnarray*}
\mathbf{J}_{1} & = & a_{-}\sum_{(ij),(rs)\in(1\pm)}\left\langle 1\right|F_{rs}^{\dagger}F_{ij}\left|1\right\rangle \left|(ij)\right\rangle \!\left\langle (rs)\right|\\
 & = & a_{-}\gamma p\left(\begin{array}{cc}
\eta_{+}^{2} & -\eta_{+}\eta_{-}\\
-\eta_{+}\eta_{-} & \eta_{-}^{2}
\end{array}\right)
\end{eqnarray*}
again rank 1, with eigenvalues $\left\{ a_{-}\gamma p,0\right\} $.
Finally, the contribution from the coherent block $\mathbf{C}$ in
the basis spanned by the basis $\left|(ij)\right\rangle $ with $(ij)\in\left\{ (0\pm),(2\pm)\right\} $
takes the form:
\begin{equation}
\mathbf{C}=a_{+}\left|C_{0}\right\rangle \!\left\langle C_{0}\right|+a_{-}\left|C_{1}\right\rangle \!\left\langle C_{1}\right|\label{eq:cohspan}
\end{equation}
because the it is diagonal in the inputs $\left\{ \left|0\right\rangle ,\left|1\right\rangle \right\} $.
In (\ref{eq:cohspan}), the vectors $\left|C_{i}\right\rangle $ collect
the amplitudes of the four Kraus operators $\left\{ F_{0\pm},F_{2\pm}\right\} $
acting on $\left|i\right\rangle $, respectively. Explicitly:
\begin{eqnarray*}
\left|C_{0}\right\rangle  & = & \left(\begin{array}{c}
\eta_{+}\sqrt{p}\\
\eta_{-}\sqrt{p}\\
\eta_{+}\mu\sqrt{1-p}\\
\eta_{-}\mu\sqrt{1-p}
\end{array}\right)\\
\left|C_{1}\right\rangle  & = & \left(\begin{array}{c}
\eta_{+}\mu\sqrt{p}\\
-\eta_{-}\mu\sqrt{p}\\
\eta_{+}\sqrt{1-p}\\
-\eta_{-}\sqrt{1-p}
\end{array}\right)
\end{eqnarray*}
and because $\mathbf{C}$ is a sum of two rank-1 matrices we have
$\mathrm{rank}\left(\mathbf{C}\right)\leq2$, i.e. $\mathbf{C}$ has
only two nonzero eigenvalues. Since $\mathbf{C}=AA^{\dagger}$ with
$A$ the 4x2 matrix:
\[
A=\left(\sqrt{a_{+}}\left|C_{0}\right\rangle ,\sqrt{a_{-}}\left|C_{1}\right\rangle \right)
\]
it also has the same eigenvalues as the Gram matrix $\mathbf{G}=A^{\dagger}A$,
\begin{equation}
\mathbf{G}=\left(\begin{array}{cc}
a_{+}\left\langle C_{0}\mid C_{0}\right\rangle  & \sqrt{a_{+}a_{-}}\left\langle C_{0}\mid C_{1}\right\rangle \\
\sqrt{a_{+}a_{-}}\left\langle C_{1}\mid C_{0}\right\rangle  & a_{-}\left\langle C_{1}\mid C_{1}\right\rangle 
\end{array}\right)
\end{equation}
here the inner products are: 
\begin{eqnarray*}
\left\langle C_{0}\mid C_{0}\right\rangle  & = & 1-(1-p)\gamma=:b_{0}\\
\left\langle C_{1}\mid C_{1}\right\rangle  & = & 1-p\gamma=:b_{1}\\
\left\langle C_{0}\mid C_{1}\right\rangle  & = & \sqrt{\left(1-\gamma\right)\left(1-\lambda\right)}
\end{eqnarray*}
\vspace{2 mm}

\begin{widetext}
then, the two eigenvalues of $\mathbf{G}$ are:
\begin{equation}
\lambda_{\pm}=\frac{1}{2}\left[\left(a_{+}b_{0}+a_{-}b_{1}\right)\pm\sqrt{\left(a_{+}b_{0}-a_{-}b_{1}\right)^{2}+4a_{+}a_{-}\left(1-\gamma\right)\left(1-\lambda\right)}\right]
\end{equation}
\end{widetext}
thus, the full complementary state has spectrum $\left\{ \lambda_{+},\lambda_{-},p\gamma a_{-},(1-p)\gamma a_{+},0,0,0,0\right\} $. 

\section{Semidefinite-programming formulation of the Holevo--Werner upper bound}
The purpose of this appendix is to provide a semidefinite programming (SDP)
formulation of the Holevo-Werner upper bound for the quantum channel
capacity and apply it to upper bound the $n$-use qubit channel representation of the dynamical map $\Phi_n$.

The Holevo--Werner bound $R_{\alpha,\Theta}\left(\mathcal{N}\right)$
is a single-letter, efficiently computable upper bound on the quantum
and private capacities of any channel $\mathcal{N}$.

It is expressed through the geometric R\'{e}nyi channel divergence:
\begin{equation}
R_{\alpha,\Theta}\left(\mathcal{N}\right)=\min_{\mathcal{M}\in\mathcal{V}_{\Theta}}D_{\alpha}\left(\mathcal{N}\parallel\mathcal{M}\right)\label{eq:optgeoren}
\end{equation}
where 
\begin{equation}
D_{\alpha}\left(\mathcal{N}\parallel\mathcal{M}\right)=\frac{1}{\alpha-1}\log\left\Vert \mathrm{Tr}_{B}\left[\mathbb{G}_{1-\alpha}\left(J_{\mathcal{N}},J_{\mathcal{M}}\right)\right]\right\Vert _{\infty}
\end{equation}
with $J_{\mathcal{N}},J_{\mathcal{M}}$ the Choi matrices of the respective
channels, $\left\Vert \cdot\right\Vert _{\infty}$ the operator norm
for an hermitian $A$: 
\begin{equation}
\left\Vert A\right\Vert _{\infty}=\lambda_{\max}\left(A\right)=\sup_{\left|\psi\right\rangle }\left\langle \psi\right|A\left|\psi\right\rangle 
\end{equation}
the largest eigenvalue of the hermitian matrix evaluated. The operator
norm also could be computed as the dual of the trace norm: 
\begin{equation}
\left\Vert A\right\Vert _{\infty}=\max_{0\preceq X\preceq I}\mathrm{Tr}\left[AX\right]
\end{equation}
where the order $A\succeq B$ means that $A-B$ is a semidefinite
matrix or equivalently that all its eigenvalues are non negative.
Then, $\mathbb{G}_{\beta}\left(\cdot,\cdot\right)$ stands for the
weighted matrix geometric mean:
\begin{equation}
\mathbb{G}_{\beta}\left(X,Y\right)=X^{1/2}\left(X^{-1/2}YX^{1/2}\right)^{\beta}X^{-1/2}
\end{equation}
and the partial trace $\mathrm{Tr}_{B}$ applies over the output system
$B$, which for qubit Choi matrices means tracing over the sub-block
content. For instance, if $J_{\mathcal{N}},J_{\mathcal{M}}$ are Choi
matrices of qubit channels, we have 4x4 block matrix: 
\begin{equation}
\mathbb{G}_{1-\alpha}\left(J_{\mathcal{N}},J_{\mathcal{M}}\right)=\left(\begin{array}{cc}
X_{00} & X_{01}\\
X_{10} & X_{11}
\end{array}\right)
\end{equation}
with every $X_{ij}$ a 2x2 matrix. Then, applying the partial trace
$\mathrm{Tr}_{B}$ leads to the 2x2 matrix:
\begin{equation}
\mathrm{Tr}_{B}\left[\mathbb{G}_{1-\alpha}\left(J_{\mathcal{N}},J_{\mathcal{M}}\right)\right]=\left(\begin{array}{cc}
\mathrm{Tr}\left(X_{00}\right) & \mathrm{Tr}\left(X_{01}\right)\\
\mathrm{Tr}\left(X_{10}\right) & \mathrm{Tr}\left(X_{11}\right)
\end{array}\right)
\end{equation}
Finally, the zero set $\mathcal{V}_{\Theta}$ for the Holevo-Werner
representation reads: 
\begin{equation}
\mathcal{V}_{\Theta}=\left\{ \mathcal{M}:\exists R_{AB}\:\mathrm{s.t.}\:R_{AB}\pm J_{\mathcal{M}}^{T_{B}}\succeq0,\mathrm{Tr}_{B}\left(R_{AB}\right)\preceq I_{A}\right\}~,
\end{equation}
where $T_{B}$ stand for the partial transpose on output system $B$.
The zero set $\mathcal{V}_{\Theta}$ collects all quantum subchannels
which cannot generate distillable entanglement, because they preserve
the positive partial transpose property. 

Additionally, note that when $J_{\mathcal{M}}$ is the Choi matrix
of a qubit subchannel it has the block structure:
\begin{equation}
J_{\mathcal{M}}=\left(\begin{array}{cc}
Y_{00} & Y_{01}\\
Y_{10} & Y_{11}
\end{array}\right)
\end{equation}
with every $Y_{ij}$ a 2x2 matrix. In this, case the partial transpose
reads: 
\begin{equation}
J_{\mathcal{M}}^{T_{B}}=\left(\begin{array}{cc}
Y_{00}^{T} & Y_{01}^{T}\\
Y_{10}^{T} & Y_{11}^{T}
\end{array}\right)
\end{equation}
where the $Y_{ij}^{T}$ are the total transpose operator applied to
every $Y_{ij}$ individually. Please note the following constraints
ensure that $\mathcal{M}$ is a valid subchannel:
\begin{eqnarray}
\mathrm{Tr}_{B}\left[J_{\mathcal{M}}\right] & \preceq & I_{A}\\
J_{\mathcal{M}} & \succeq & 0
\end{eqnarray}
however one needs to include only $J_{\mathcal{M}}\succeq0$, because
the other constraint is already implied by the other conditions:
\begin{eqnarray*}
\mathrm{Tr}_{B}\left[J_{\mathcal{M}}\right] & = & \mathrm{Tr}_{B}\left[J_{\mathcal{M}}^{T_{B}}\right]\\
 & \preceq & \mathrm{Tr}_{B}\left(R_{AB}\right)\\
 & \preceq & I_{A}.
\end{eqnarray*}

In reference \cite{FangFawzi2021}, the authors show in Proposition
19 that an optimization like the right hand side of (\ref{eq:optgeoren})
admits a semidefinite representation for $\alpha=1+2^{-l}$ with $l\in\mathbb{N}.$
Applying the Proposition 19 to $R_{\alpha,\Theta}\left(\mathcal{N}\right)$
we obtain: 
\begin{eqnarray*}
 & R_{\alpha,\Theta}\left(\mathcal{N}\right)=2^{l}\log\min y\\
s.t. & R\pm N_{0}^{T_{B}}\succeq0 & I_{A}-\mathrm{Tr}_{B}\left(R\right)\succeq0\\
 & N_{0}\succeq0\\
 & \left(\begin{array}{cc}
J_{\mathcal{N}} & N_{i}\\
N_{i} & N_{i-1}
\end{array}\right)\succeq0 & i=1,\ldots,l\\
 & \left(\begin{array}{cc}
M & J_{\mathcal{N}}\\
J_{\mathcal{N}} & N_{l}
\end{array}\right)\succeq0 & yI_{A}-\mathrm{Tr}_{B}\left(M\right)\succeq0
\end{eqnarray*}
with $y\in\mathbb{R}$ and all $R,M,N_{i}$ for $i=0,...,l$ hermitian
matrices of the same dimension of the Choi matrix $J_{\mathcal{N}}$.
Note, the above SDP has the form of a dual problem.

Now, for the specific case of the $n$-use qubit channel representation of the dynamical map $\Phi_n$, the corresponding
Choi matrix is:
\begin{widetext}
\begin{equation}
J_{\mathcal{E}_{\Phi_n}}\!\!\!=\!\!\left(\!\!\!\begin{array}{cccc}
1-\gamma_n(1-p_n) & 0 & 0 & \sqrt{(1-\gamma_n)(1-\lambda_n)}\\
0 & \gamma_n(1-p_n) & 0 & 0\\
0 & 0 & \gamma_n p_n & 0\\
\sqrt{(1-\gamma_n)(1-\lambda_n)} & 0 & 0 & 1-\gamma_n p_n
\end{array}\!\!\right)
\end{equation}
with parameters $\lambda_n,p_n,\gamma_n\in\left[0,1\right]$. 
\end{widetext}

Numerically, the above SDP is implemented in Python using
CVXPY~\cite{Diamond2016} and solved with the Clarabel interior-point
solver~\cite{Goulart2024}, converging to optimality across all plotted points up
to $l=10$ within duality-gap and feasibility tolerances of $5\times 10^{-7}$ and
$10^{-7}$ respectively. Whenever the analytical PPT condition
$(1-\gamma_n)(1-\lambda_n)\le\gamma_n^2 p_n(1-p_n)$ is satisfied
($n\ge n_{\mathrm{EB}}$), the channel is entanglement-breaking and we set
$\mathrm{HW}(n,l)=0$ directly without invoking the solver. For the analytical
bounds, probabilities entering the binary and Shannon entropies are clipped
below at $10^{-15}$ to avoid logarithmic singularities, and $\mathrm{G}(n,l)$ is undefined whenever $\mathrm{BN}(n)-\mathrm{LB}(n)\le 10^{-12}$.

\section{Optimal second-use certification threshold}
\label{app:second_use_threshold}

In this appendix, we establish that the three-site chain maximizes the relative timing-error tolerance certified by our lower bound on the second use. We consider the equal-readout-time protocol and write
\begin{equation}
\delta_{\mathrm{LB}}^{(N)}(2)
=
\sup\left\{
\delta\in[0,1]:
\mathrm{LB}^{(N)}(2,\delta)>0
\right\}.
\end{equation}
The proof constructs an upper bound on $\mathrm{LB}^{(N)}(2,\delta)$ that is nonincreasing with $N$. For every $N\ge4$, this upper bound vanishes before the three-site lower bound loses positivity.

For convenience, define
\begin{equation}
r=\cos^2\!\left(\frac{\pi\delta}{2}\right),
\qquad
s_N=1-\gamma_2^{(N)}=r^{N-1}.
\end{equation}
The coherence factor on the second use is
\begin{equation}
a_N
\equiv A_1^{(N)}
=\sqrt{1-\lambda_2^{(N)}}
=r^{N-1}+(1-r)^{N-1}.
\end{equation}
This expression follows from single-excitation probability conservation: the first memory factor equals the probability that the excitation occupies either endpoint after the first transfer.

After removing the known output phase, the effective channel acts as
\begin{equation}
\Phi_2^{(N)}
\begin{pmatrix}
1-q & c\\
c^* & q
\end{pmatrix}
=
\begin{pmatrix}
1-s_Nq-\beta_N & \sqrt{s_N}\,a_Nc\\
\sqrt{s_N}\,a_Nc^* & s_Nq+\beta_N
\end{pmatrix},
\end{equation}
where
\begin{equation}
\beta_N
=
\bigl(1-p_2^{(N)}\bigr)(1-s_N).
\end{equation}
The diagonal input states entering the lower bound are denoted by
\begin{equation}
\rho_q=\operatorname{diag}(1-q,q),
\qquad
q=\frac{1-z}{2}.
\end{equation}

We first introduce the auxiliary damping--dephasing channel
\begin{equation}
\mathcal{F}_{s,a}
\begin{pmatrix}
1-q & c\\
c^* & q
\end{pmatrix}
=
\begin{pmatrix}
1-sq & \sqrt{s}\,ac\\
\sqrt{s}\,ac^* & sq
\end{pmatrix},
\qquad
0\le s,a\le1.
\end{equation}
Writing $\mathcal{X}(\rho)=X\rho X$, the effective channel admits the decomposition
\begin{equation}
\begin{aligned}
\Phi_2^{(N)}
={}&
p_2^{(N)}\mathcal{F}_{s_N,a_N}
\\
&+
\bigl(1-p_2^{(N)}\bigr)
\mathcal{X}\circ\mathcal{F}_{s_N,a_N}\circ\mathcal{X}.
\end{aligned}
\end{equation}

For a fixed input state, coherent information is convex in the channel. Indeed, it equals the negative conditional entropy of the output obtained from a purification of that input, and conditional entropy is concave. Unitary invariance therefore gives
\begin{equation}
\begin{aligned}
I_c\!\left(\rho_q,\Phi_2^{(N)}\right)
&\le
p_2^{(N)}
I_c\!\left(\rho_q,\mathcal{F}_{s_N,a_N}\right)
\\
&\quad+
\bigl(1-p_2^{(N)}\bigr)
I_c\!\left(\rho_{1-q},\mathcal{F}_{s_N,a_N}\right).
\end{aligned}
\end{equation}
Consequently, defining
\begin{equation}
U_N(\delta)
=
\max_{q\in[0,1]}
I_c\!\left(\rho_q,\mathcal{F}_{s_N,a_N}\right),
\end{equation}
we obtain
\begin{equation}
0\le
\mathrm{LB}^{(N)}(2,\delta)
\le U_N(\delta).
\end{equation}

The auxiliary bound is nonincreasing with chain length. At fixed $\delta$,
\begin{equation}
s_{N+1}=rs_N,
\qquad
a_{N+1}\le a_N.
\end{equation}
Moreover, the auxiliary channels satisfy the composition rule
\begin{equation}
\mathcal{F}_{s_{N+1},a_{N+1}}
=
\mathcal{F}_{r,\,a_{N+1}/a_N}
\circ
\mathcal{F}_{s_N,a_N}.
\end{equation}
Since $0\le a_{N+1}/a_N\le1$, the additional map is a valid quantum channel. Data processing of coherent information then implies
\begin{equation}
U_{N+1}(\delta)\le U_N(\delta).
\end{equation}

The damping--dephasing channel $\mathcal{F}_{s,a}$ is antidegradable when \cite{Siddhu2025}:
\begin{equation}
1-s\ge\frac{a^2}{1+a^2},
\qquad\text{equivalently}\qquad
s(1+a^2)\le1.
\end{equation}
In this regime its coherent information is nonpositive for every input. Since pure inputs yield zero coherent information, $U_N(\delta)=0$ whenever this condition holds.

For $N=4$, the condition becomes
\begin{equation}
r^3\left[1+\left(r^3+(1-r)^3\right)^2\right]\le1.
\end{equation}
On $r\in[1/2,1]$, the left-hand side is strictly increasing with $r$. For $r\le1/2$, it is bounded above by $1/4$. Thus, if $r_4$ denotes the unique solution of
\begin{equation}
r_4^3
\left[1+\left(r_4^3+(1-r_4)^3\right)^2\right]
=1,
\end{equation}
then
\begin{equation}
U_4(\delta)=0
\qquad\text{for}\qquad
\delta\ge\bar{\delta}_4,
\end{equation}
where
\begin{equation}
\bar{\delta}_4
=
\frac{2}{\pi}\arccos\sqrt{r_4}
\simeq0.22512052.
\end{equation}
Combining this result with monotonicity in $N$ yields
\begin{equation}
\delta_{\mathrm{LB}}^{(N)}(2)
\le\bar{\delta}_4
\simeq22.5121\%,
\qquad N\ge4.
\end{equation}

We now evaluate the three-site lower bound. Suppressing the chain-length superscript, its parameters are
\begin{equation}
s=r^2,
\qquad
a=r^2+(1-r)^2,
\qquad
\beta=2r^2(1-r)^2.
\end{equation}
The coherent information for a diagonal input is
\begin{equation}
\begin{aligned}
I_3(q,\delta)
={}&h(sq+\beta)
\\
&-
H\!\left(
\ell_+,\ell_-,
(1-s-\beta)q,\,
\beta(1-q)
\right),
\end{aligned}
\end{equation}
where $h$ and $H$ denote the binary and Shannon entropies in bits, respectively, and
\begin{equation}
\begin{aligned}
\ell_\pm
=
\frac12\Bigl[
&(1-q)(1-\beta)+q(s+\beta)
\\
&\pm
\sqrt{
\left[(1-q)(1-\beta)-q(s+\beta)\right]^2
+
4q(1-q)sa^2
}
\Bigr].
\end{aligned}
\end{equation}
Accordingly,
\begin{equation}
\mathrm{LB}^{(3)}(2,\delta)
=
\max_{q\in[0,1]}I_3(q,\delta).
\end{equation}

A direct witness already establishes that the three-site threshold exceeds the upper limit obtained for every longer chain. Choose
\begin{equation}
r_0=\frac{22}{25},
\qquad
q_0=\frac{11}{50},
\end{equation}
and
\begin{equation}
\delta_0
=
\frac{2}{\pi}\arccos\sqrt{\frac{22}{25}}
\simeq22.5199\%.
\end{equation}
Substitution into the three-site expression gives
\begin{equation}
I_3(q_0,\delta_0)
\simeq4.39376\times10^{-3}>0.
\end{equation}
Hence,
\begin{equation}
\delta_{\mathrm{LB}}^{(N)}(2)
\le\bar{\delta}_4
<
\delta_0
\le
\delta_{\mathrm{LB}}^{(3)}(2),
\qquad N\ge4.
\end{equation}
This proves the optimality of the three-site chain independently of the numerical determination of its certification boundary.

For completeness, the three-site lower bound is nonincreasing with timing error over the interval containing this boundary. Let $\Psi_r$ denote the three-site channel parametrized by $r$. For
\begin{equation}
\frac12\le r_2\le r_1\le1,
\end{equation}
define
\begin{equation}
\kappa=\frac{r_2^2}{r_1^2},
\qquad
b=\frac{a(r_2)}{a(r_1)},
\qquad
\eta=\beta(r_2)-\kappa\beta(r_1).
\end{equation}
These parameters satisfy
\begin{equation}
0\le b\le1,
\qquad
0\le\eta\le1-\kappa.
\end{equation}
The lower bound on $\eta$ follows from
\begin{equation}
\eta
=
2r_2^2
\left[(1-r_2)^2-(1-r_1)^2\right]
\ge0.
\end{equation}
For the upper bound, introduce
\begin{equation}
g(r)=\frac{1-\beta(r)}{r^2}
=\frac1{r^2}-2(1-r)^2.
\end{equation}
Since
\begin{equation}
g'(r)=-\frac2{r^3}+4(1-r)<0,
\end{equation}
we have
\begin{equation}
1-\kappa-\eta
=
r_2^2\bigl[g(r_2)-g(r_1)\bigr]
\ge0.
\end{equation}

Therefore, the transformation
\begin{equation}
q\longmapsto\kappa q+\eta,
\qquad
c\longmapsto\sqrt{\kappa}\,bc
\end{equation}
defines a valid generalized-amplitude-damping channel composed with dephasing, denoted by $\mathcal{T}$, and
\begin{equation}
\Psi_{r_2}=\mathcal{T}\circ\Psi_{r_1}.
\end{equation}
Data processing establishes that the three-site lower bound cannot increase as $\delta$ increases on $0\le\delta\le1/2$. For $\delta\ge1/2$, $s=r^2\le1/4$, and damping-induced antidegradability already guarantees a vanishing lower bound.

Finally, numerical maximization of the explicit three-site expression locates the certification boundary. At this boundary, the interior maximum satisfies
\begin{equation}
I_3(q_*,\delta_*)=0,
\qquad
\frac{\partial I_3}{\partial q}(q_*,\delta_*)=0.
\end{equation}
Optimizing over the full interval $q\in[0,1]$ gives
\begin{equation}
q_*\simeq0.21880174,
\qquad
\delta_*\simeq0.2262425809.
\end{equation}
Thus,
\begin{equation}
\max_{N\ge3}\delta_{\mathrm{LB}}^{(N)}(2)
=
\delta_{\mathrm{LB}}^{(3)}(2)
\simeq22.6243\%.
\end{equation}
The maximal relative timing-error tolerance certified by our lower bound on the second use therefore lies just below $22.63\%$ and is attained by the three-site chain.

\bibliography{n_use.bib}
\end{document}